\documentclass[%
 aip,
 floatfix,
 reprint,%
 author-year,%
]{revtex4-1}

\usepackage{graphicx}
\usepackage{amsmath,amssymb,amsfonts,mathrsfs,bm}
\usepackage{mathtools}
\usepackage[labelformat=simple]{subcaption}
\usepackage{siunitx}

\makeatletter
\newcommand*{\balancecolsandclearpage}{%
  \close@column@grid
  \cleardoublepage
  \twocolumngrid
}
\makeatother

\usepackage{hyperref}
 \hypersetup{
     colorlinks=true,
     linkcolor=blue,
     filecolor=blue,
     citecolor=blue,      
     urlcolor=blue
}
\usepackage{breakurl}

\allowdisplaybreaks

\begin{document}

\title{Data-driven rheological characterization of stress buildup and relaxation in thermal greases}

\author{Pranay P. Nagrani}
\affiliation{School of Mechanical Engineering, Purdue University, West Lafayette, Indiana 47907, USA}

\author{Ritwik V. Kulkarni}
\affiliation{School of Mechanical Engineering, Purdue University, West Lafayette, Indiana 47907, USA}

\author{Parth U. Kelkar}
\affiliation{School of Materials Engineering, Purdue University, West Lafayette, Indiana 47907, USA}

\author{Ria D. Corder}
\affiliation{School of Mechanical Engineering, Purdue University, West Lafayette, Indiana 47907, USA}
\affiliation{School of Materials Engineering, Purdue University, West Lafayette, Indiana 47907, USA}

\author{Kendra A. Erk}
\affiliation{School of Materials Engineering, Purdue University, West Lafayette, Indiana 47907, USA}

\author{Amy M. Marconnet}
\affiliation{School of Mechanical Engineering, Purdue University, West Lafayette, Indiana 47907, USA}

\author{Ivan C. Christov}
\thanks{Author to whom correspondence should be addressed}
\email{christov@purdue.edu}
\affiliation{School of Mechanical Engineering, Purdue University, West Lafayette, Indiana 47907, USA}

\date{\today}

\begin{abstract}
Thermal greases, often used as thermal interface materials, are complex paste-like mixtures composed of a base polymer in which dense metallic (or ceramic) filler particles are dispersed to improve the heat transfer properties of the material. They have complex rheological properties that impact the performance of the thermal interface material over its lifetime. We perform rheological experiments on thermal greases and observe both stress relaxation and stress buildup regimes. This time-dependent rheological behavior of such complex fluid-like materials is not captured by steady shear-thinning models often used to describe these materials. We find that thixo-elasto-visco-plastic (TEVP) and nonlinear-elasto-visco-plastic (NEVP) constitutive models characterize the observed stress relaxation and buildup regimes respectively. Specifically, we use the models within a data-driven approach based on physics-informed neural networks (PINNs). PINNs are used to solve the inverse problem of determining the rheological model parameters from the dynamic response in experiments. This training data is generated by startup flow experiments at different (constant) shear rates using a shear rheometer. We validate the ``learned'' models by comparing their predicted shear stress evolution to experiments under shear rates not used in the training datasets. We further validate the learned TEVP model by solving a forward problem numerically to determine the shear stress evolution for an input step-strain profile. Meanwhile, the NEVP model is further validated by comparison to a steady Herschel--Bulkley fit of the material's flow curve. 
\end{abstract}


\maketitle

\section{Introduction}
\label{sec::Intro}

Miniaturization of electronic packages \citep{Bird1995ApproachesMiniaturization,Henry1950NewMiniaturization,Iwai2021ImpactAfter} makes the dissipation of heat to the environment of paramount importance to the reliability of microsystems. Typically, a thermal grease \citep{Zhou2020RecentMaterials} of high thermal conductivity that can conform to the surface and fill gaps is employed to reduce the otherwise high contact thermal resistance between two solid substrates. Thermal greases, being soft, viscoelastic materials exhibiting both solid-like and liquid-like rheological behavior, can keep their shape at rest, yet they flow under a sufficiently strong applied force. Thus, their rheology is complex and perhaps even time-dependent \citep{Coussot2007RheophysicsApproaches}. Furthermore, thermal greases degrade over time through the processes of pump out and dry out \citep{Chiu2001AnApplications,Nnebe2008Drainage-inducedGreases,Gowda2003DesignMicroelectronics,Hayden2020}. The former causes void formation, which degrades the heat dissipation properties of the assembly and can thus cause an overshoot of allowable junction level temperatures. It is hypothesized that the degradation of thermal greases can be understood by first characterizing their rheology behavior \citep{Prasher2004ThermalMaterial,Prasher2003ThermalMaterials,Prasher2006ThermalDirections,Lin2009RheologicalPastes}, which provides the motivation for the present work.


Thermal greases are often modeled as shear-thinning fluids, and their rheological characterization is limited to simple shear flows at steady state. The most common such models are the Herschel--Bulkley (HB) and the Bingham models \citep{Prasher2002RheologicalExperimental,Prasher2002RheologicalModeling}. The latter incorporates a yield stress, while the former, in addition, also accounts for shear thinning in a steady flow. Within an electronic package, thermal greases are present between two solid substrates to enhance heat transfer. The thickness of the thermal grease layer (on the order of a few microns) sandwiched between the two substrates is known as the bond line thickness. Earlier work \citep{Prasher2003ThermalMaterials,Prasher2006ThermalDirections,Prasher2002RheologicalModeling} focused on developing a semi-empirical model for the bond line thickness of a thermal grease comprised of filler particles within a silicone oil. Typically thermal greases are squeezed to a fixed pressure, and hence the final bond line thickness depends on the rheological properties of the grease, such as its effective viscosity and yield stress. For high filler particle concentrations, the semi-empirical rheological model is modified to incorporate the effect of a percolation threshold (i.e., a chain of filler particles contacting one another across the bond line thickness resulting in enhanced heat transfer)  \citep{Prasher2005RheologyMaterials}. The viscoelastic behavior of thermal greases was studied by \citet{Lin2009RheologicalPastes}, who used a polyester matrix-based thermal paste with varying volume fractions of solid particles (carbon black, fumed alumina, and nanoclay). They observed Bingham plastic behavior at high concentrations of particles. Meanwhile, a fluid-like behavior was observed when no solid particles were present. Clearly, thermal greases are complex soft materials, and their rheological behavior and thermal performance are strongly influenced by the particles and other ``modifier'' materials used to enhance the wetting between polymer and particles. To this end, \citet{Feger2005MixingPastes} evaluated the influence of different mixing processes on thermal paste rheology. They concluded that strongly-sheared mixing yields more homogeneous thermal pastes, leading to a smaller bond line thickness and, hence, reduced junction level temperatures. Beyond these studies, no clear picture has emerged of the rheological behavior of thermal greases under prescribed shear (whether fixed, stepped, or varying). In this work, we provide a perspective on these behaviors using experiments and fundamental rheological models within a data-driven framework.


Physics-informed neural networks (PINNs) are a class of machine learning techniques that simultaneously solve a direct and an inverse problem \citep{Karniadakis2021Physics-informedLearning}. Specifically, PINNs can be used to train a neural network to solve a system of ordinary or partial differential equations and simultaneously identify unknown physical parameters in these governing equations \citep{Raissi2019Physics-informedEquations}. The success of PINNs has been established in many contexts within the mechanics of fluids and solids \citep[e.g.,][]{Lu2023Physics-informedFlow,Cai2021Physics-informedReview,Henkes2022PhysicsMicromechanics,Haghighat2021AMechanics,Cai2021Physics-InformedProblems}. A recent application of PINNs to rheology led to the development of the so-called  RhINNs (Rheology-Informed Neural Networks) \citep{Mahmoudabadbozchelou2021Data-drivenFramework,Mahmoudabadbozchelou2021Rheology-InformedFluids,Saadat2022Data-drivenRhINNs,Mahmoudabadbozchelou2022Nn-PINNs:Modeling}. RhINNs have been shown to successfully calibrate time-dependent rheological models of complex fluids using only a small number of experimental measurements. In particular, \citet{Mahmoudabadbozchelou2021Rheology-InformedFluids} constructed RhINNs based on three constitutive models: a thixo-visco-plastic, a thixo-elasto-visco-plastic (TEVP), and an iso-kinematic hardening model. The direct problem was solved to predict the shear stress profile of a complex fluid subjected to different flow protocols: startup, flow-hysteresis, small amplitude oscillatory shear, and large amplitude oscillatory shear. Simultaneously, the RhINNs solved the inverse problem of calibrating these models by inferring the unknown models' parameters, hence characterizing the rheology of the complex fluids considered. Further, RhINNs have been used to select the best-fit steady-state constitutive model for a given complex fluid, using the flow curve (i.e., the variation of viscosity with shear rate) as an input to the RhINN \citep{Saadat2022Data-drivenRhINNs}. In this work, following the more standard terminology, we will refer to all PINN-based methods as `PINNs,' whether the physics incorporated are rheological models or not.

One major shortcoming of the previous rheological characterization efforts of thermal greases is that the models and experiments neither address stress relaxation of a thermal grease due to step-shear nor the shear-rate-dependent or even time-dependent rheological behavior (thixotropy). The goal of the present work is to combine physics-informed machine learning with state-of-the-art fundamental ideas from rheology to accurately calibrate general models of thermal grease flow behavior, which we hypothesize will eventually enable us to better predict thermal grease degradation and failure modes.  As \citet{Mahmoudabadbozchelou2022DigitalNetworks} argued, this kind type of data-driven approach can be considered a ``digital rheometer twin,'' able to characterize the ``hidden physics'' of complex fluids from standard rheological measurement. To this end, we leverage the open-source platform DeepXDE \citep{Lu2021DeepXDE:Equations} to develop PINNs to characterize the rheology of two Dow thermal greases with substantially different flow behaviors. The specific greases considered are DOWSIL TC-5622 and DOWSIL TC-5550, which exhibit distinct flow behaviors. The former exhibits stress relaxation, while the latter exhibits stress buildup. Hence, for the physics incorporated within the PINNs, we use a transient TEVP model \citep{Mahmoudabadbozchelou2021Rheology-InformedFluids,Jamali2017MicrostructuralFluid,DeSouzaMendes2011ThixotropicFluids}, as well as the insightful model for viscoplasticity proposed by \citet{Kamani2021UnificationFluids} to unify the behavior of rheologically nonlinear fluids above and below their yield stress.

This paper is organized as follows. In Sec.~\ref{sec:CM}, we introduce the constitutive models used to characterize the rheology of the thermal greases under consideration. The PINN construction is detailed in Sec.~\ref{sec:RhINN}. Next, Sec.~\ref{sec::expt_method} describes the protocol for rheological experiments. The corresponding rheological characterizations, and the demonstrations of the different flow regimes (stress buildup and relaxation), are carried out in Sec.~\ref{sec::RAndD}. Finally, Sec.~\ref{sec::Conclusion} summarizes our results. 

\section{Modeling methodology}
\label{sec::model_methods}

In this section, we describe the constitutive models used to characterize the rheology of the DOWSIL TC-5622 and DOWSIL TC-5550 thermally conductive compounds under consideration. The only prior information about these thermal greases is found in their data sheets \citep{Dow2017DOWSILSheet,Dow2022DOWSILSheet}. In particular, the exact composition (filler particle type, volume fraction, particle diameter, etc.) is a trade secret and unknown to the authors. Hence, to select appropriate constitutive models to characterize DOWSIL TC-5622 and DOWSIL TC-5550, we first conducted rheological experiments on two thermal greases and observed that they both exhibit time-dependent rheological behavior, but not the same one. Our initial observations on the greases suggested that they possess a yield stress. In addition to the yield stress and their complex microstructure, the greases seemed to exhibit elastic resistance and recovery. Next, we examined recent reviews on such viscoplastic fluids \citep{Coussot2007RheophysicsApproaches,Balmforth2014YieldingMechanics,Frigaard2019SimpleFluids} to seek a suitable rheological model for the observed behaviors. Eventually, we arrived at the works of \cite{Mahmoudabadbozchelou2021Rheology-InformedFluids} and \cite{Kamani2021UnificationFluids} and found that the models discussed therein aim to capture the rheological behaviors of interest to us. 

On the basis of these observations and prior literature on the shear-thinning behavior of thermal greases, two suitable constitutive models are introduced in Sec.~\ref{sec:CM}. Then, we will demonstrate the models' utility by introducing them into a data-driven, PINN-based framework. The training of the models using the experimental rheological data is discussed in Sec.~\ref{sec:RhINN}.

\subsection{Constitutive models}
\label{sec:CM}

From a rheological perspective, thermal greases exhibit viscous resistance and a time-dependent behavior beyond their yield stress. In particular, stress relaxation is often observed in thermal grease degradation during pump-out \citep{Chiu2001AnApplications,Nnebe2008Drainage-inducedGreases,Gowda2003DesignMicroelectronics,Hayden2020}. Arguably, the simplest transient rheology model with these features is the elasto-visco-plastic (EVP) model. However, this model neither accounts for shear thinning nor captures stress relaxation physics during step-strain experiments. These shortcomings of the EVP model are overcome by accounting for \emph{thixotropy} \citep{Barnes1997ThixotropyaReview,Ewoldt2017MappingBehavior,Larson2019AModeling}, resulting in the TEVP model. Indeed, when subjected to shear, the thermal grease microstructure evolves, which results in a time-dependent shear-thinning response. Based on this fact and our initial experimental observations, we hypothesize that DOWSIL TC-5622 exhibits thixotropy. We believe this behavior arises from the presence of particles within the polymer matrix, which adjust their position when subjected to shear. Below, we show that the TEVP model can be used to characterize this thixotropic flow behavior of thermal greases exhibiting stress relaxation.

There are a variety of thixotropic rheological models proposed in the literature \citep{Jamali2017MicrostructuralFluid,DeSouzaMendes2011ThixotropicFluids,Dullaert2006AThixotropy,Larson2019AModeling}.  We use the version of the TEVP model defined by \citet{Mahmoudabadbozchelou2021Rheology-InformedFluids}, which takes the form:
\begin{subequations}
\begin{align}
    \frac{d\sigma}{dt} &= \frac{G}{\eta_s+\eta_p} \left\{ -\sigma(t) + \sigma_y\lambda(t) + [\eta_s+\eta_p\lambda(t)] \dot \gamma(t) \right\},\\
    \frac{d\lambda}{dt} &= k_{+} [1-\lambda(t)] - k_{-}\dot{\gamma}(t)\lambda(t),
\end{align}
\label{eq:TEVP-dim}%
\end{subequations}
where $\sigma$ is the shear stress, $G$ is an elastic shear modulus, and $\eta_s$ and $\eta_p$ are the solvent and plastic viscosities, respectively; $\sigma_y$ is the yield stress, $\dot{\gamma}$ is the applied shear rate, and $t$ is time. Further, $\lambda$ is the so-called structure parameter describing the degree of solid-like behavior. In this simple model, $\lambda=1$ represents a structured (or solid-like) state of the material, while $\lambda=0$ represents an unstructured (or fluid-like) state of the material's microstructure. Here, $k_{+}$ and $k_{-}$ are termed the buildup and breakage coefficients, respectively. 
As seen from Eq.~\eqref{eq:TEVP-dim}, the TEVP model consists of two ordinary differential equations (ODEs). The first ODE captures the stress evolution (elasto-visco-plastic behavior) given the imposed shear. The second ODE describes the evolution of microstructure. The two coupled ODEs characterize the complex material's time-dependent rheological response (thixotropy). 

The shear stress during a startup flow can be measured in a rheometer. The structure parameter, however, is not directly measurable as it depends on the microstructure of the thermal grease, which cannot be directly characterized in a rheometer. Hence, using the TEVP model within a physics-informed machine learning framework allows us to infer (and understand) the evolution of the microstructure when the thermal grease is subjected to constant shear rates. As described in Sec.~\ref{sec::RAndD} below, we calibrate the unknown model parameters ($G$, $\eta_p$, $\eta_s$, $\sigma_y$, $k_{+}$, and $k_{-}$) from experimental measurements of the stress response by constructing and training a PINN.
 
Meanwhile, our initial experimental observations regarding the DOWSIL TC-5550 thermal grease suggested a significant stress buildup regime at low shear rates. This behavior cannot be captured by the TEVP model. Thus, we propose to characterize the rheology of the DOWSIL TC-5550 using the nonlinear-elasto-visco-plastic (NEVP) model proposed by \citet{Kamani2021UnificationFluids}, which aims to unify the time-dependent rheological behavior of complex materials before and after yielding using a single equation. The model, assuming a constant shear rate (i.e., $d{\dot{\gamma}}/dt = 0$), takes the form
\begin{equation}
      \label{eq:unification_dim}
      \frac{d\sigma}{dt}  = \frac{G\dot\gamma(t)}{\sigma_y+k\dot\gamma(t)^{n}+\eta_s\dot\gamma(t)} \left\{\sigma_y + k \dot \gamma(t)^n -\sigma(t)  \right\},
\end{equation}
where $k$ is the consistency index, and $n$ is the power-law exponent from an HB fit ($\sigma = \sigma(\dot{\gamma}) = \sigma_y + k \dot\gamma^{n}$) of the steady flow curve ($d\sigma/dt=d\dot\gamma/dt=0$); the remaining parameters are as in Eq.~\eqref{eq:TEVP-dim}. In this model, the unknown parameters to be calibrated are  $G$, $\eta_s$, $\sigma_y$, $k$, and $n$.
Under steady shear rate conditions and at low shear rates (here, $\dot\gamma \le 0.1~\si{\per\second}$), the NEVP model will be used below to characterize the ``elastic regime'' of stress buildup, which is relevant within an electronic package subjected to thermo-mechanical stresses (analogous to nonzero shear rates). 

\begin{figure*}
    \centering\includegraphics[width=0.85\linewidth]{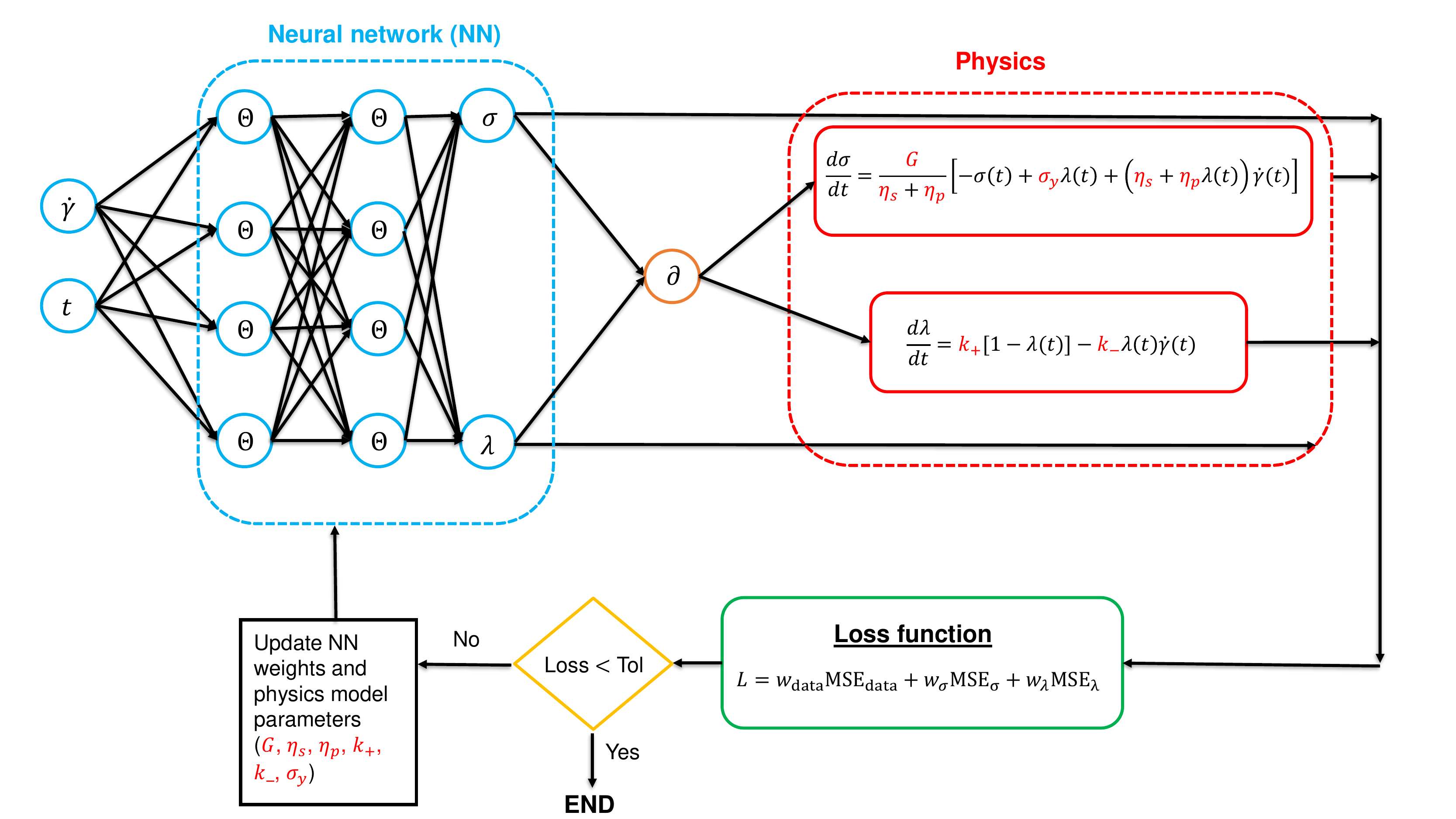}
    \caption{Flowchart of the PINN architecture and training process used in this work. A deep neural network is used to evaluate quantities in the TEVP constitutive model, whose residual is minimized via the physics-informed loss function. Here, $\Theta$ represents the neurons of the neural network. The unknown model parameters of the TEVP constitutive model are highlighted in red. In the training process, the total loss, comprising of the mean squared error (MSE) of the constitutive model residual and the training data MSE is minimized. ``Tol" is a user-defined tolerance for each model, as described in the main text. Note that this flowchart would be modified for the NEVP model, for which there is only a single constitutive equation and a single output, recall Eq.~\eqref{eq:unification_dim}.}
    \label{fig::PINN_FlowChart}
\end{figure*}

\subsection{PINN construction}
\label{sec:RhINN}

The PINNs employed in this work to characterize thermal grease rheology use the architecture shown in Fig.~\ref{fig::PINN_FlowChart}. Specifically, a deep neural network of densely connected neurons, each associated with weights and biases, is employed. The network has two inputs, $t$ and $\dot \gamma$, and two outputs, $\sigma$ and $\lambda$, as the schematic in Fig.~\ref{fig::PINN_FlowChart} represents the TEVP model~\eqref{eq:TEVP-dim}. When the NEVP model is employed, there is only one output $\sigma$ and one physics equation, namely Eq.~\eqref{eq:unification_dim}. The outputs are then used to compute the residual of the physics ODEs (i.e., the constitutive models) by leveraging automatic differentiation \citep{Margossian2019AImplementation}. The workflow is implemented in DeepXDE \citep{Lu2021DeepXDE:Equations}.

\begin{widetext}
The overall loss function (for the TEVP model) is the sum of the mean-squared error (MSE) of matching the training data ($\mathrm{MSE}_{\mathrm{data}}$) and the MSEs of satisfying the constitutive equations ($\mathrm{MSE}_{\sigma}$ and $\mathrm{MSE}_{\lambda}$). 
Using arbitrary weights $w$ for each MSE, the total loss $L$ is written as
\begin{subequations}
\begin{align} 
    L &= w_\mathrm{data} \mathrm{MSE_\mathrm{data}} + w_{\sigma} \mathrm{MSE}_{\sigma} + w_{\lambda} \mathrm{MSE}_{\lambda},\label{eq:LossDef}\\
    \mathrm{MSE_\mathrm{data}} &= \frac{1}{M} \sum_{i=1}^{M} | \sigma_{\mathrm{experiment}}(t_i) - \sigma(t_i) |^2, \\
    \mathrm{MSE}_{\sigma} &= \frac{1}{N} \sum_{i=1}^{N} \left| \left.\frac{d\sigma}{dt}\right|_{t=t_i} - \frac{G}{\eta_s+\eta_p} \left\{ -\sigma(t_i) + \sigma_y\lambda(t_i) + [\eta_s+\eta_p\lambda(t_i)] \dot \gamma(t_i) \right\} \right|^2,\\
    \mathrm{MSE}_{\lambda} &= \frac{1}{N} \sum_{i=1}^{N} \left| \left.\frac{d\lambda}{dt}\right|_{t=t_i} - k_{+} [1-\lambda(t_i)] + k_{-}\dot{\gamma}(t_i)\lambda(t_i)\right|^2,
\end{align}
\label{eq:OverallLoss}%
\end{subequations}
where $M$ is the number of training data points; $N$ is the number of domain points sampled in the input space; $\{\sigma_{\mathrm{experiment}}(t_i)\}$ are the experimental data points obtained by performing startup flow experiments in a rheometer (see Sec.~\ref{sec::expt_method}) and $\{\sigma(t_i)\}$ are the corresponding shear stress values predicted by the PINN. 
\end{widetext}
For the NEVP model, the loss function consists of only one term for the constitutive model's residual loss (along with training data loss).

In training a PINN, the objective is to minimize the overall loss (i.e., the sum of the training data loss from experiments and the constitutive equation residual loss) by simultaneously optimizing the neural network weights \emph{and} the physical model's unknown parameters. To this end, we chose the machine learning hyper-parameters based on the choices justified in related recent studies \citep{Lu2023Physics-informedFlow,Chen2020Physics-informedMetamaterials,Cai2021Physics-InformedProblems,Li_ENERGY} and experimentation. Specifically, the Adam optimizer \citep{Adam}, a $\tanh$ activation function, and Glorot normal initialization were used. The neural network architecture consisted of $10$ neurons per layer and a depth of $4$ layers. We conducted a parametric study with different neural network depths ($3$, $4$, and $5$) and different numbers of neurons per layer ($10$, $30$, and $50$) to conclude that the final training loss is of the same order, $10^{-4}$, in all cases; hence it is relatively independent of the architecture. In addition, since $\lambda$ from the TEVP model is physically restricted to the interval $[0,1]$, we apply a sigmoid function to the $\lambda$ output of the NN. Further, $w_{\sigma}=1$, $w_{\lambda}=1$, and $w_\mathrm{data}=10$ were used to train the TEVP PINN, while $w_{\sigma}=1$ and $w_\mathrm{data}=100$ were used to train the NEVP PINN. The weights were calibrated to avoid over-fitting and, at the same time, provide similar orders of magnitude for the losses due to the constitutive model residual and the data residual (recall Eq.~\eqref{eq:LossDef}). 

Finally, to avoid under- or over-fitting, we normalize each input and output variable, call its discrete values $\xi_i$, over all  data points $i=1,2,\hdots$, via 
\begin{equation}
    \xi_i \mapsto \frac{\xi_i - \min_j \xi_j}{\max_j \xi_j - \min_j \xi_j}.
\end{equation} 
For example, $\xi$ is any of $\sigma$, $t$ or $\dot \gamma$.  The constitutive equations~\eqref{eq:TEVP-dim} and \eqref{eq:unification_dim} are also made dimensionless, using the normalization introduced, to be consistent with the training data. Finally, the neural network in the TEVP PINN undergoes training for approximately $1.5 \times 10^{6}$ epochs (or iterations) at a learning rate of $1 \times 10^{-3}$ to achieve an overall loss of $L\le 6 \times 10^{-4}$. Meanwhile, the neural network in the NEVP PINN  undergoes training for $10^{6}$ epochs, at a learning rate of $5 \times 10^{-4}$, to achieve an overall loss of $L\le 10^{-3}$. (We used a looser threshold for $L$ for the NEVP model to prevent over-fitting to the noise present in the shear stress profiles at low $\dot\gamma$.) Finally, the PINN training takes approximately $5$ hours on an $8$-core machine with $16$ GB of RAM.

\section{Experimental methodology}
\label{sec::expt_method}

This section illustrates the experimental protocol we adopted to generate training data for the DOWSIL TC-5622 and DOWSIL TC-5550 thermal greases (Dow Chemical Company). The experiments were performed using an Anton Paar TwinDrive MCR702 rheometer with a single gap concentric cylinder setup having a bob diameter $10.0~\si{\milli\meter}$, bob length $14.9~\si{\milli\meter}$ and measuring gap $0.422~\si{\milli\meter}$ by applying a constant shear rate to a material initially at rest (i.e., a startup flow protocol). Three repeatable experimental datasets were obtained for each thermal grease. One dataset was used to generate the results in this work, and the other two datasets are available online (see ``Data Availability'' statement).

\subsection{DOWSIL TC-5622}

\begin{figure}
    \centering\includegraphics[width=\linewidth]{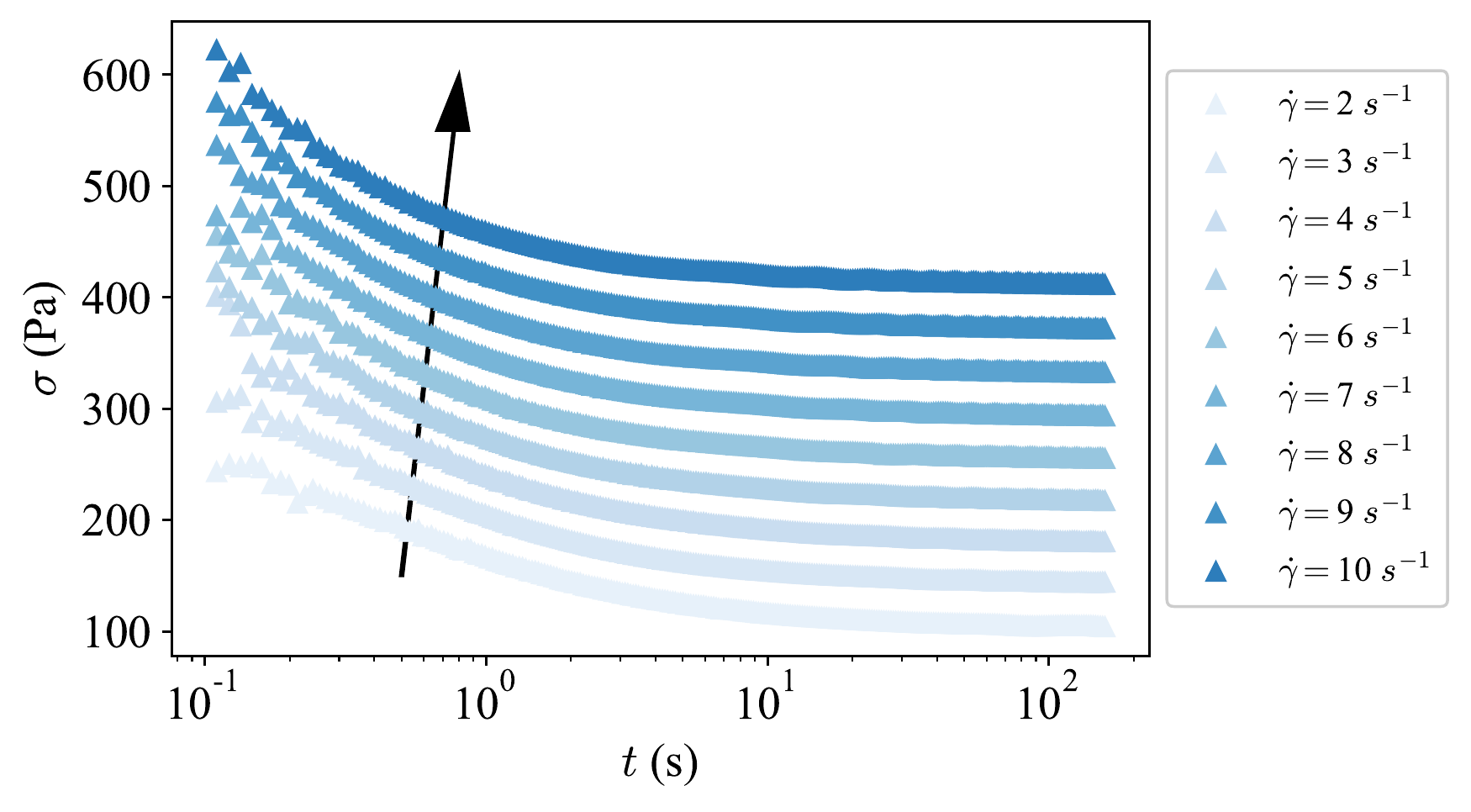}
    \caption{Training data obtained from rheological experiments by subjecting DOWSIL TC-5622 to startup flow at steady shear rates ranging from $\dot{\gamma} = 2~\si{\per\second}$ to $\dot{\gamma} = 10~\si{\per\second}$ for approximately $180~\si{s}$. The arrow indicates the direction of increasing $\dot{\gamma}$. Only the data up to $10~\si{\second}$ is used for training as it captures all the stress relaxation physics.}
    \label{fig::DOW5622_expt}
\end{figure}

We hypothesize that the DOWSIL TC-5622 thermal grease is thixotropic in nature. To substantiate this hypothesis, we performed standard experiments \citep{Barnes1997ThixotropyaReview,DeSouzaMendes2011ThixotropicFluids,Jamali2017MicrostructuralFluid} to characterize this thermal grease's stress relaxation. Specifically, we subjected the thermal grease to a startup flow protocol. To erase memory effects, the material was subjected to preshear of $2~\si{\per\second}$ for $180~\si{s}$, followed by a shear-free rest period of  $180~\si{s}$. Next, we subjected the material to a startup flow protocol by suddenly imposing a fixed, nonzero shear rate for approximately $180~\si{s}$ to observe the stress relaxation profile. 
The shear rates used were $2~\si{\per\second}$ to $10~\si{\per\second}$, and data was obtained using log sampling to obtain more data points at early times, as shown in Fig.~\ref{fig::DOW5622_expt}. The original data was truncated at $t = 10~\si{\second}$ for every $\dot{\gamma}$ as, by this time, shear stresses reached an approximate steady state. The rheology and stress relaxation of DOWSIL TC-5622 is characterized by employing the TEVP constitutive model~\eqref{eq:TEVP-dim} (i.e., learn/determine $G$, $\eta_s$, $k_{+}$, $k_{-}$, $\eta_p$, $\sigma_y$) and the startup flow  experiment data from Fig.~\ref{fig::DOW5622_expt}.

We note that, for low shear rates $\dot{\gamma} < 2~\si{\per\second}$, some of our rheological data exhibited nonmonotone behavior at early times, which is not shown. We do not believe that the measured noisy, nonmonotone early-time data for low shear rates can be trusted, and indeed there is evidence that steady, homogeneous flows of complex yield stress materials may not exist below a critical shear rate due to shear banding \citep{Ovarlez2009PhenomenologyFluids}. Therefore, for this study, we restrict our shear rates to $2~\si{\per\second} \leq \dot{\gamma} \leq 10~\si{\per\second}$ for the rheological characterization of DOWSIL TC-5622 within the relaxation regime. 

A similar experimental flow protocol was used to generate data for a step-strain shear rate profile to be used to validate the learned model parameters in  Sec.~\ref{sec:relax}. In this case, we first subject the DOWSIL TC-5622 to preshear of $2~\si{\per\second}$ for $180~\si{s}$, followed by a shear-free rest period of $180~\si{s}$. Next, we subject DOWSIL TC-5622 to the step-strain shear rate profile given by Eq.~\eqref{eq:step-strain}.

\subsection{DOWSIL TC-5550}

\begin{figure}[t]
    \centering\includegraphics[width=\linewidth]{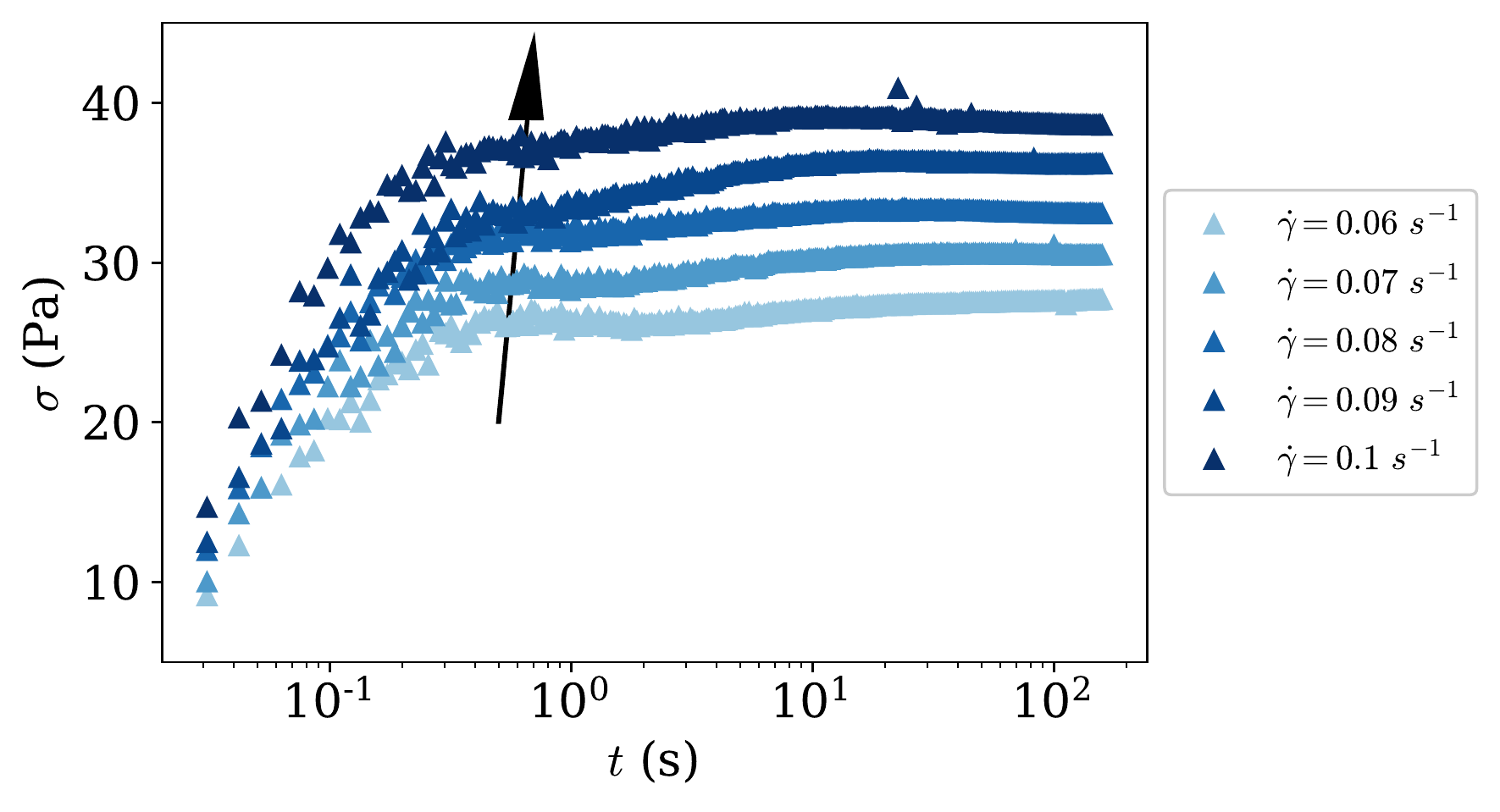}
    \caption{Training data obtained from rheological experiments by subjecting DOWSIL TC-5550 to startup flow  at steady shear rates of $0.06~\si{\per\second}$ to $0.1~\si{\per\second}$ for approximately $180~\si{s}$. The arrow indicates the direction of increasing $\dot{\gamma}$. It should be noted that data up to $10~\si{\second}$ is used for training as it captures all the stress buildup physics.}
    \label{fig::DOWSIL TC-5550_expt}
\end{figure}

To characterize the stress buildup (the so-called `elastic regime') and rheology of the DOWSIL TC-5550 thermal grease, startup flow experiments were conducted at low shear rates (i.e., from  $\dot{\gamma}=0.06~\si{\per\second}$ to $\dot{\gamma}=0.1~\si{\per\second}$), as shown in Fig.~\ref{fig::DOWSIL TC-5550_expt}. The shear stress increases in time but then saturates, indicating that the material response is reaching a steady state. Log sampling was used to obtain more data points in the high-shear-stress-gradient region (i.e., for $t<1~\si{\second}$). 

For larger shear rates, $\dot{\gamma} > 0.1~\si{\per\second}$, the rheological data exhibited nonmonotone behavior at early times, which is not shown. We do not believe that this noisy, nonmonotone data can be trusted; hence, we restrict our shear rates to $0.06~\si{\per\second} \leq \dot{\gamma} \leq 0.1~\si{\per\second}$ for the rheological characterization of DOWSIL TC-5550 within the buildup regime. The experimental protocol was similar to that for DOWSIL TC-5622: we subjected the thermal grease to preshear of $\dot\gamma=4~\si{\per\second}$ followed by a rest period of $240~\si{\second}$, after which the startup flow experiments at a suddenly-imposed, nonzero shear rate were performed. Since DOWSIL TC-5550 is more viscous than DOWSIL TC-5622, a larger preshear rate was employed to fluidize the material (and hence erase memory effects) before subjecting it to startup flow. We used data up to $t=10~\si{s}$, at which time we again observed that an approximate steady state was established, as seen in Fig.~\ref{fig::DOWSIL TC-5550_expt}. The rheology and stress buildup in the DOWSIL TC-5550 thermal grease was characterized  by employing the NEVP constitutive model~\eqref{eq:unification_dim} (i.e., learning $G$, $\eta_s$, $\sigma_y$, $k$, and $n$) and the startup flow experiment data from Fig.~\ref{fig::DOWSIL TC-5550_expt}.

As a cross-check for the learned model parameters, we performed a flow-curve experiment by ramping shear rates from $15~\si{\per\second}$ to $0.05~\si{\per\second}$ (after subjecting the DOWSIL TC-5550 thermal grease to preshear and a rest period, as described above). The flow curves allow us to determine a steady HB model. Each data point (corresponding to a different shear rate) was sampled at a $10~\si{\second}$ interval, which was deemed sufficient to reach a steady state (see Fig.~\ref{fig::DOWSIL TC-5550_expt}). Finally, the values for $k$ and $n$ from the HB fit and the value for $\sigma_y$ from extrapolating the steady flow curve were compared to the values learned by the PINN to ascertain the validity and consistency of the data-driven machine learning approach proposed. 

\section{Results and discussion}
\label{sec::RAndD}

In this section, we discuss the PINN predictions for the startup flow profile and the learned model parameters for the two thermal greases under consideration. In Sec.~\ref{sec:relax} and Sec.~\ref{sec:elastic}, we validate the prediction for each thermal grease. Each case represents a different flow regime and a corresponding rheological model. We posit that each regime is of relevance to a different process in an industrial device (such as an electronics package). For example, the range $\dot\gamma\in[2,10]~\si{\per\second}$ for DOWSIL TC-5622 leads to stress relaxation behavior, which is relevant to the pumpout (or outward flow) behavior within an electronic package. Meanwhile, the lower range, $\dot\gamma\in[0.06,0.1]~\si{\per\second}$ for DOWSIL TC-5550, is relevant to the stress buildup observed before degradation begins.

\subsection{Characterizing the stress relaxation regime}
\label{sec:relax}
For DOWSIL TC-5622, we used the startup flow data from Fig.~\ref{fig::DOW5622_expt} to train a PINN based on the TEVP constitutive model. As training data, we used startup flow data from all shear rates, except $\dot\gamma = 7~\si{\per\second}$.  Then, we used the $\dot\gamma = 7~\si{\per\second}$ data to test the PINN's prediction. The underlying PINN architecture minimizes the total loss~\eqref{eq:OverallLoss}. Although, in this formulation, we do not give any additional constraints in the form of initial or boundary conditions, as these are not known \textit{a priori}, providing such data significantly improves the PINNs' ability to extrapolate, as discussed in Appendix~\ref{app:PINN_extrapolation}.

\begin{table}
    \setlength{\tabcolsep}{5pt}
    \begin{tabular}{lll}
      \hline
      \hline
      Parameter  & Value \\
      \hline
      $G$ & $196.1$ \si{\pascal}\\
      $\eta_s$ & $39.4$ \si{\pascal\second} \\
      $k_{+}$ & $0.06$ \si{\per \second}\\
      $k_{-}$  & $0.06$ \\
      $\sigma_y$ & $31.1$ \si{\pascal} \\
      $\eta_p$ & $29.1$ \si{\pascal\second} \\
      \hline
      \hline
    \end{tabular}
    \caption{Learned values of the TEVP rheological model's parameters, as introduced in the constitutive equations~\eqref{eq:TEVP-dim}, for DOWSIL TC-5622.}
    \label{tab::dow5622}
\end{table} 

\begin{figure}
    \centering
    \begin{subfigure}[t]{\linewidth}
      \includegraphics[width=\linewidth]{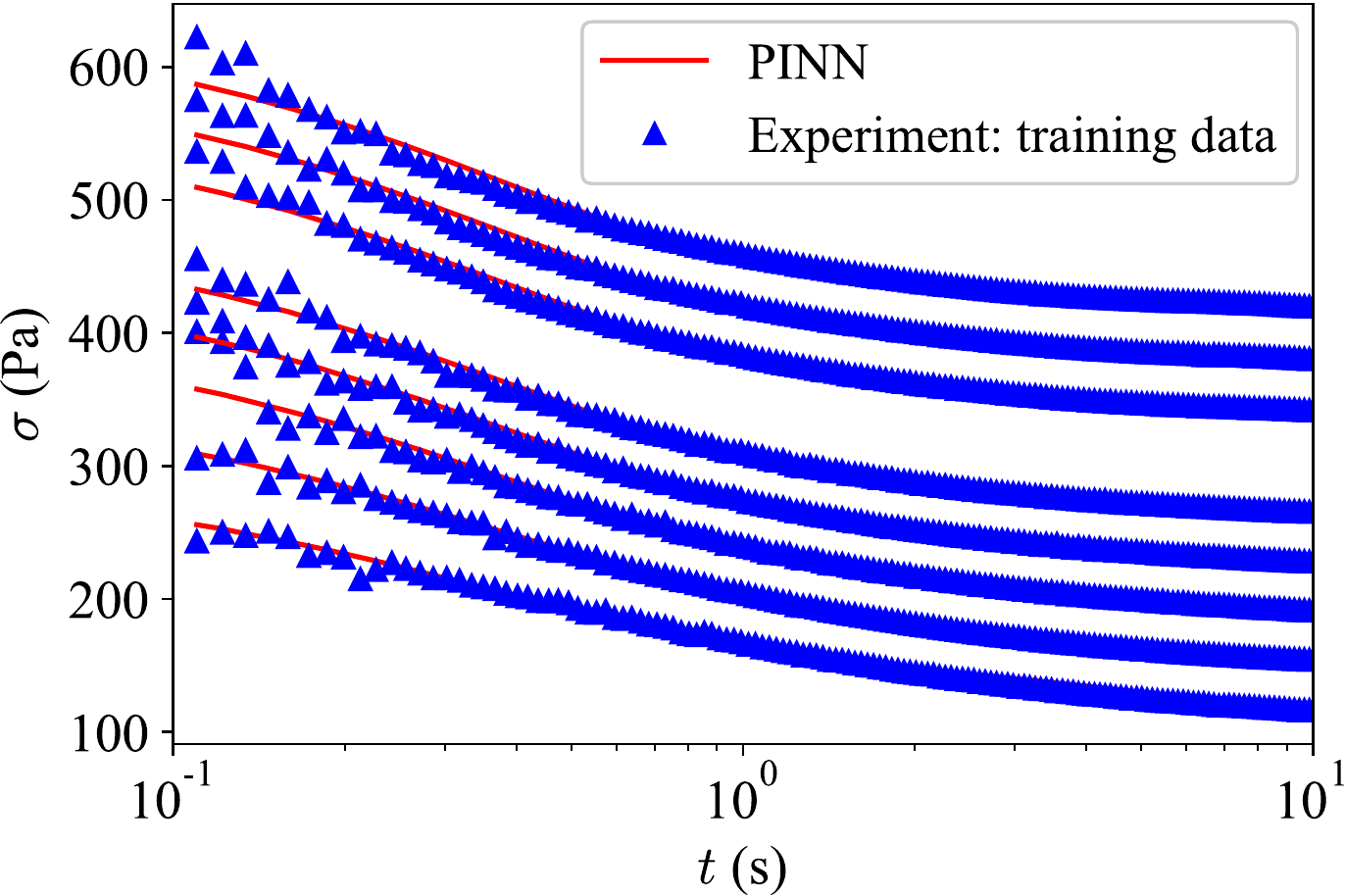}  
      \caption{}
      \label{fig:5622_training}
    \end{subfigure}
    \begin{subfigure}[t]{\linewidth}
      \includegraphics[width=\linewidth]{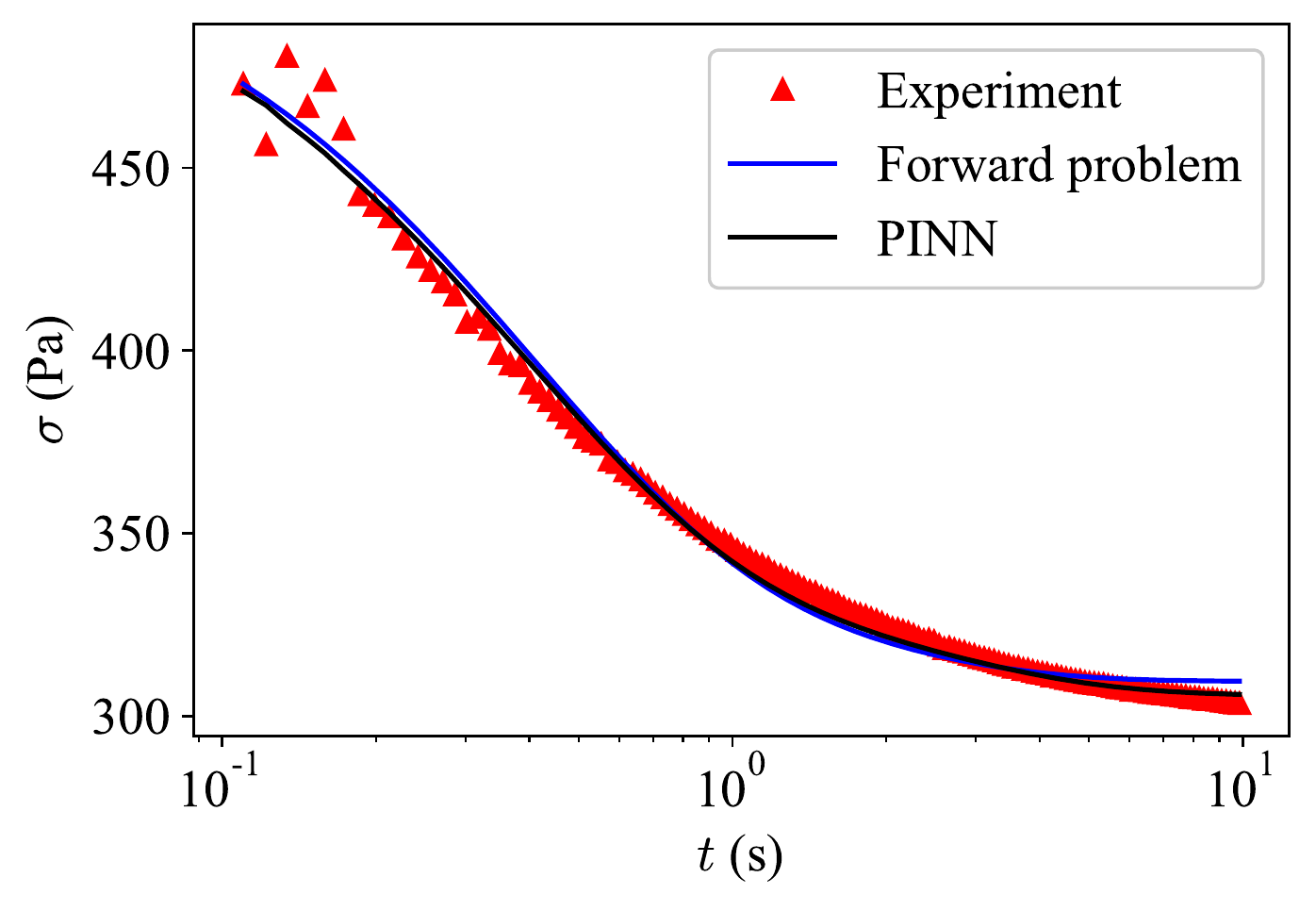}  
      \caption{}
      \label{fig:5622_test}
    \end{subfigure}
    \caption{Comparison between PINN predictions and the experimental data, for (a) training data and (b) unseen data (for startup flow with $\dot\gamma = 7~\si{\per\second}$), for the DOWSIL TC-5622 thermal grease. The unknown model parameters learned by the PINN are summarized in Table~\ref{tab::dow5622}.}
    \label{fig::dow5622}
\end{figure}

As seen from Fig.~\ref{fig:5622_training}, the training data at all shear rates are well predicted by the PINN, which is expected because this set of data is given as input for training, and the deep neural network is a good function approximator. Further, Fig.~\ref{fig:5622_test} shows that after the PINN is trained, its performance on unseen test data (startup flow with $\dot\gamma = 7~\si{\per\second}$) is satisfactory; specifically, it tracks the experimental stress relaxation profile accurately.
Table~\ref{tab::dow5622} lists the learned model parameters for DOWSIL TC-5622. An average value of each unknown model parameter is reported, obtained by retraining the PINN, excluding shear stress data for different $\dot\gamma$ each time, and relearning the model parameters. Using these values, we solve Eqs.~\eqref{eq:TEVP-dim} as a forward problem (using Python SciPy's \texttt{odeint} subroutine; \citealp{SciPy}). As can be seen in Fig.~\ref{fig:5622_test}, there is good agreement between the forward problem solution, the PINN's prediction, and the experiment data. In addition, we independently obtain $\sigma_y=29.5~\si{\pascal}$ from a flow curve experiment (see Appendix~\ref{app::dow5622-flow-curve}), which is in agreement with the value reported in Table~\ref{tab::dow5622}. 

For further validation of the learned model parameters, we also numerically solve the forward problem for the TEVP
model~\eqref{eq:TEVP-dim} (in Python using SciPy's \texttt{odeint} subroutine; \citealp{SciPy}) for a stepped-shear profile. Specifically, the learned values of the unknown model parameters (summarized in Table~\ref{tab::dow5622}) were used to predict the stress profile for an input step-strain shear rate profile:
\begin{equation}\label{eq:step-strain}
\dot\gamma(t) = 
\begin{cases}
1~\si{\per\second}, & 0~\si{\second} < t \le 158.25~\si{\second},\\
0.1~\si{\per\second}, & 158.25~\si{\second} < t \le 180.042~\si{\second},\\
1~\si{\per\second}, & 180.042~\si{\second} < t \le  338.292~\si{\second}.
\end{cases}
\end{equation}

\begin{figure}
    \centering
    \includegraphics[width=\linewidth]{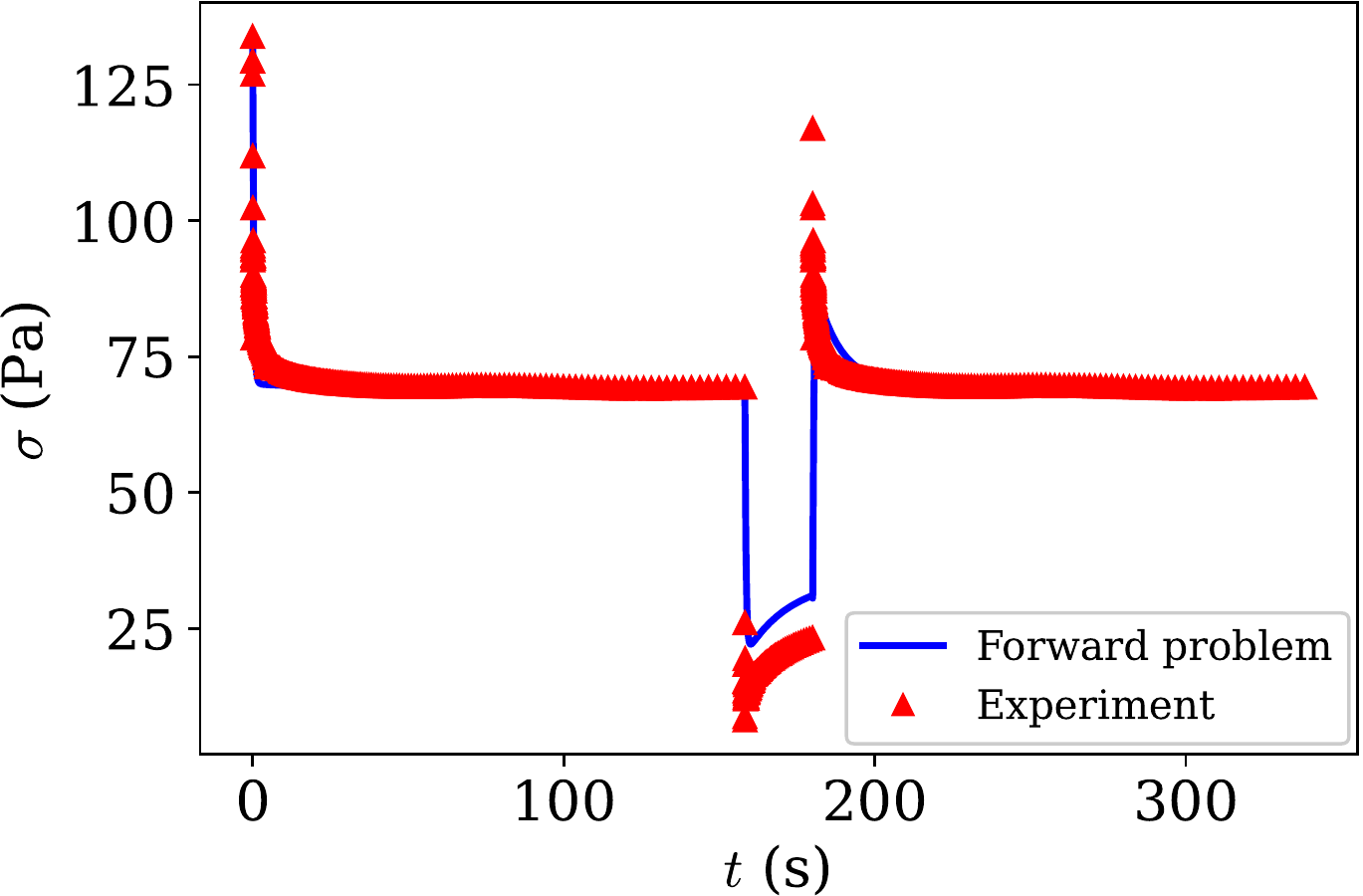}
    \caption{Validation of the calibrated TEVP rheology of DOWSIL TC-5622 in the stress relaxation regime under the input step-strain shear rate profile from Eq.~\eqref{eq:step-strain}. The constitutive equations~\eqref{eq:TEVP-dim} were solved numerically as a direct problem using the model parameters values in Table~\ref{tab::dow5622}. The experimentally measured stress response is shown as symbols.}
    \label{fig::DOW-5622-val}
\end{figure}

The shear stress obtained by solving the direct problem by forward integration is compared to the experimental data in Fig.~\ref{fig::DOW-5622-val}. Initial conditions (for $\sigma$ and $\lambda$) are required to solve the direct problem by forward integration. We chose the values $\sigma(t=0)=133.79~\si{\pascal}$ (obtained from the step-strain experiment used for validation) and $\lambda(t=0)=0.5$ (i.e., assuming partial breakdown of microstructure). A sensitivity analysis was performed on the initial condition of the microstructure parameter $\lambda(t=0)$, and we found that any value $\lambda(t=0)\in[0.5,1]$ yields similar results. The direct problem captures the stress relaxation profile accurately followed by a dip in stress observed due to step-down from $\dot\gamma = 1~\si{\per\second}$ to $\dot\gamma = 0.1~\si{\per\second}$ and finally the step-up from $\dot\gamma = 0.1~\si{\per\second}$ to $\dot\gamma = 1~\si{\per\second}$. We believe that the few extreme data points at which the experimental data and the forward problem profile do not agree are likely because of the very sudden change in $\dot\gamma$ during step-strain, which might not be well described by a one-dimensional rheological model. Nevertheless, these are only a few isolated points. Therefore, we are confident in the learned unknown model parameters. In turn, we have demonstrated a robust method of characterizing the rheology (within the relaxation flow regime) of DOWSIL TC-5622 thermal grease using a PINN to perform data-driven calibration of the rheological model. 

To test the sensitivity of the PINN formulation with respect to the initial guesses for the model parameters, we trained PINNs for three different sets of initial guesses. We found that learned model parameters are rather insensitive to the initial guesses, as shown in  Appendix~\ref{app::parspace}. Further, to demonstrate the PINN's ability to infer the microstructure evolution, $\lambda(t)$, without being given any data for it, in Appendix~\ref{app:PINN_synthetic_data} we generated synthetic data and retrained the PINN on it, showing reasonable agreement between the expected and PINN-generated $\lambda(t)$ profiles.

\subsection{Characterizing the stress buildup  regime}
\label{sec:elastic}
The DOWSIL TC-5550 thermal grease exhibits stress buildup (the so-called ``elastic'' regime) before yielding. This rheological response was characterized using the NEVP model~\eqref{eq:unification_dim}. As described above, the PINN was used to estimate the unknown model parameters in a data-driven fashion. The data from Fig.~\ref{fig::DOWSIL TC-5550_expt} (excluding $\dot\gamma = 0.08~\si{\per\second}$) was used for training.

\begin{figure}
    \centering
    \begin{subfigure}[t]{\linewidth}
      \includegraphics[width=\linewidth]{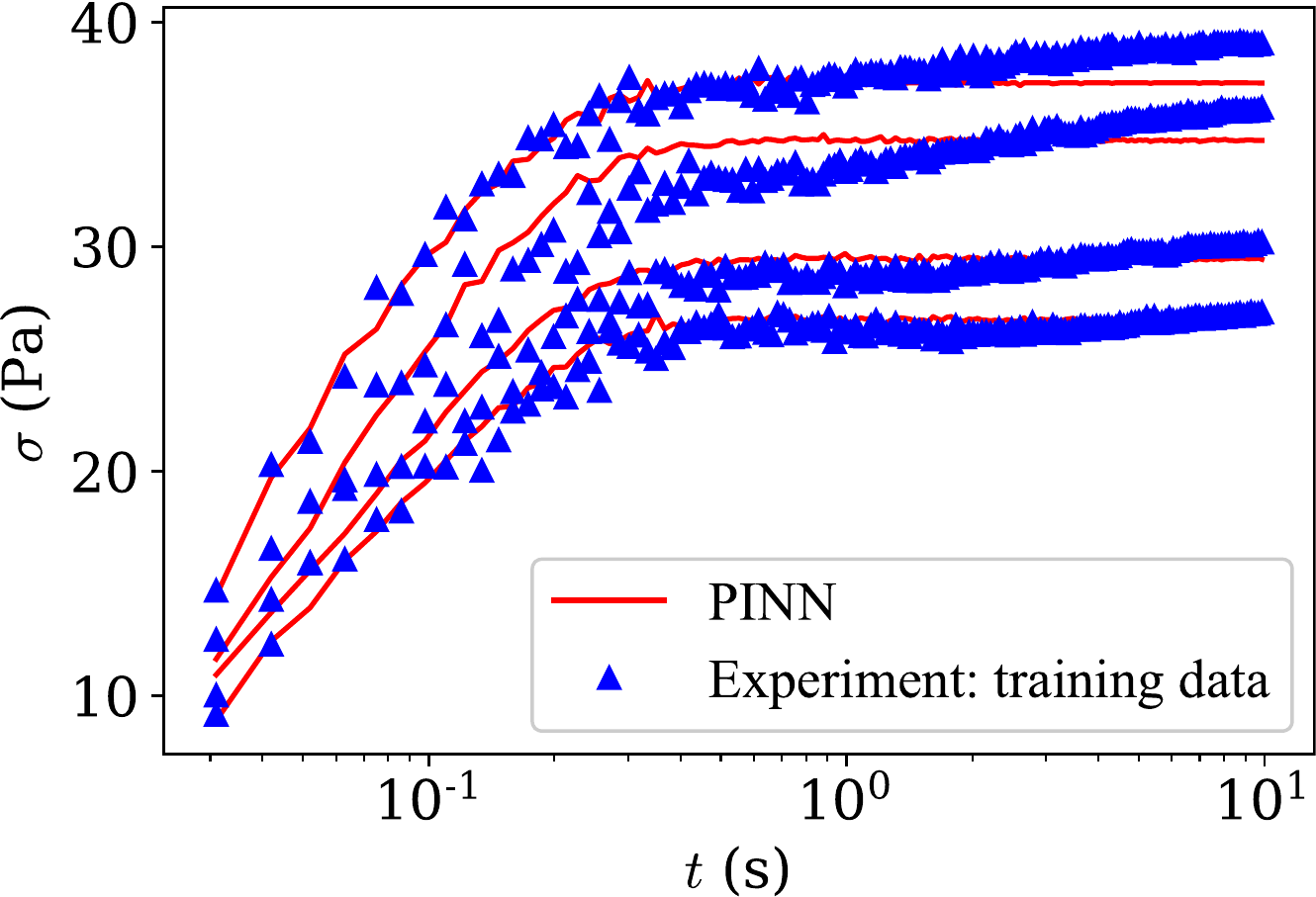}  
      \caption{}
      \label{fig:1_training}
    \end{subfigure}
    \begin{subfigure}[t]{\linewidth}
      \includegraphics[width=\linewidth]{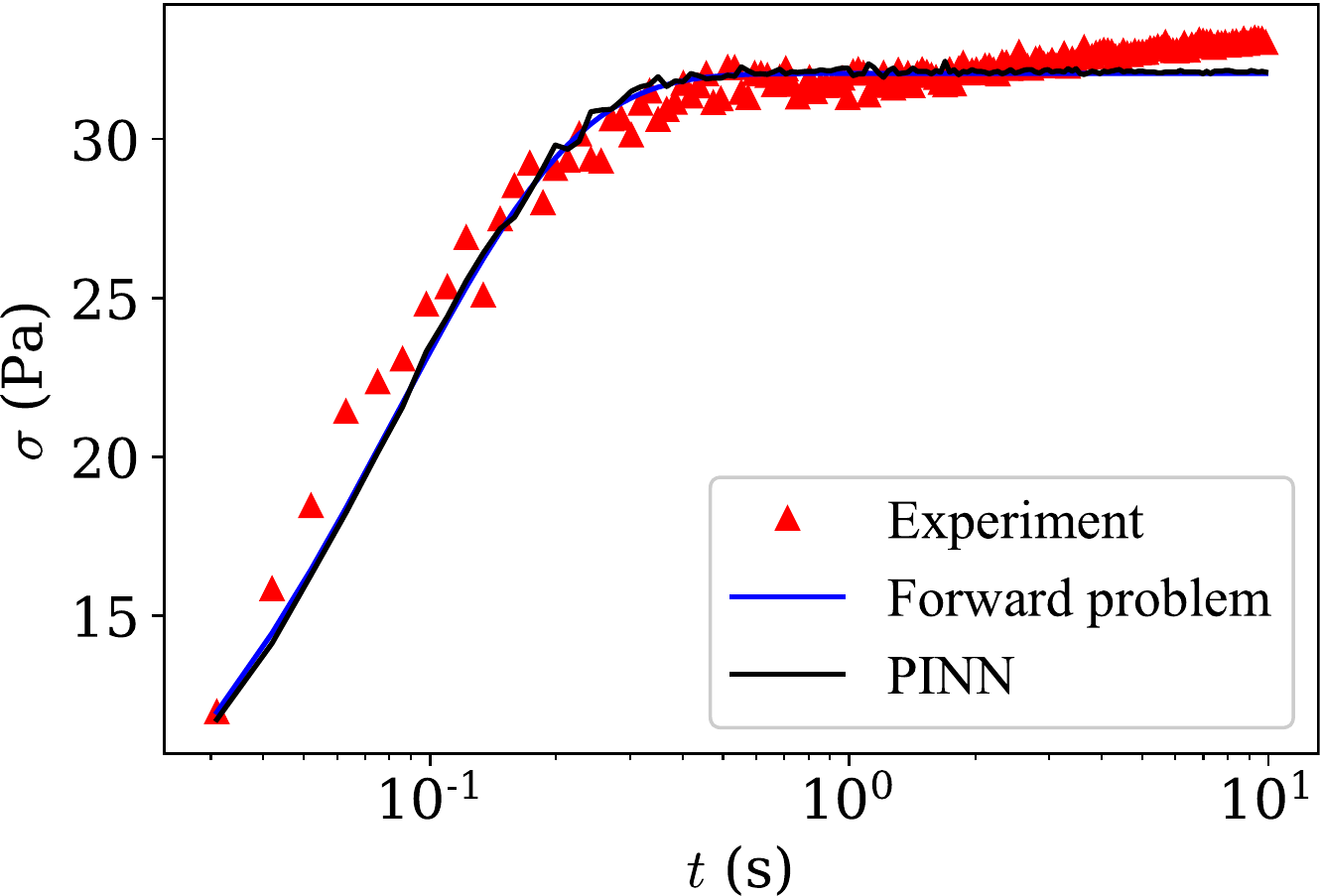}  
      \caption{}
      \label{fig:1_test}
    \end{subfigure}
    \caption{Comparison between the PINN predictions and the experimental data, for (a) training data and (b) unseen startup flow data (for $\dot\gamma = 0.08~\si{\per\second}$), for the DOWSIL TC-5550 thermal grease. The unknown model parameters learned by the PINN are summarized in Table~\ref{tab::dow1}.}
    \label{fig::dow1}
\end{figure}

As seen from Fig.~\ref{fig:1_training}, the PINN accurately predicts the input training data. This agreement is expected as the deep neural network is a good function approximator. More importantly, the PINN predicts the preyielding behavior of the DOWSIL TC-5550 thermal grease at the unseen shear rate $\dot\gamma = 0.08~\si{\per\second}$, as shown in Fig.~\ref{fig:1_test}. The learned model parameters are listed in Table~\ref{tab::dow1}. We use these values to solve the forward problem for the NEVP model~\eqref{eq:unification_dim} (using Python SciPy's \texttt{odeint} subroutine; \citealp{SciPy}). We observe good agreement between the forward problem, the PINN prediction, and the experiment data. No additional constraints in the form of initial or boundary conditions were imposed for the training. An average value of each unknown model parameter is reported, obtained by retraining the PINN, excluding shear stress data for different $\dot\gamma$ each time, and relearning the model parameters. Note that the typical $\eta_p$ values estimated for DOWSIL TC-5550 (Table~\ref{tab::dow1}) is a factor of ten larger than for DOWSIL TC-5622 (Table~\ref{tab::dow5622}). This observation further highlights how different the rheophysics are for these two materials. The DOWSIL TC-5550 grease resists the imposed shear much more strongly, leading to a well-defined stress buildup regime, which is well captured by the NEVP model.

\begin{table}
    \centering
    \setlength{\tabcolsep}{10pt}
    \begin{tabular}{ll}
      \hline
      \hline
      Parameter  & Value \\
      \hline
      $G$ & $8972.0$ \si{\pascal}\\
      $\eta_s$ & $363.0$ \si{\pascal\second} \\
      $k$  & $214.2$ \si{\pascal \second \tothe{n}} \\
      $n$  & $0.85$ \\
      $\sigma_y$ & $6.9$ \si{\pascal} \\
      \hline
      \hline      
    \end{tabular}
    \caption{Learned values of the NEVP rheological model's parameters, as introduced in the constitutive equation~\eqref{eq:unification_dim}, for DOWSIL TC-5550.}
    \label{tab::dow1}
\end{table} 

\begin{figure}
    \centering
    \includegraphics[width=\linewidth]{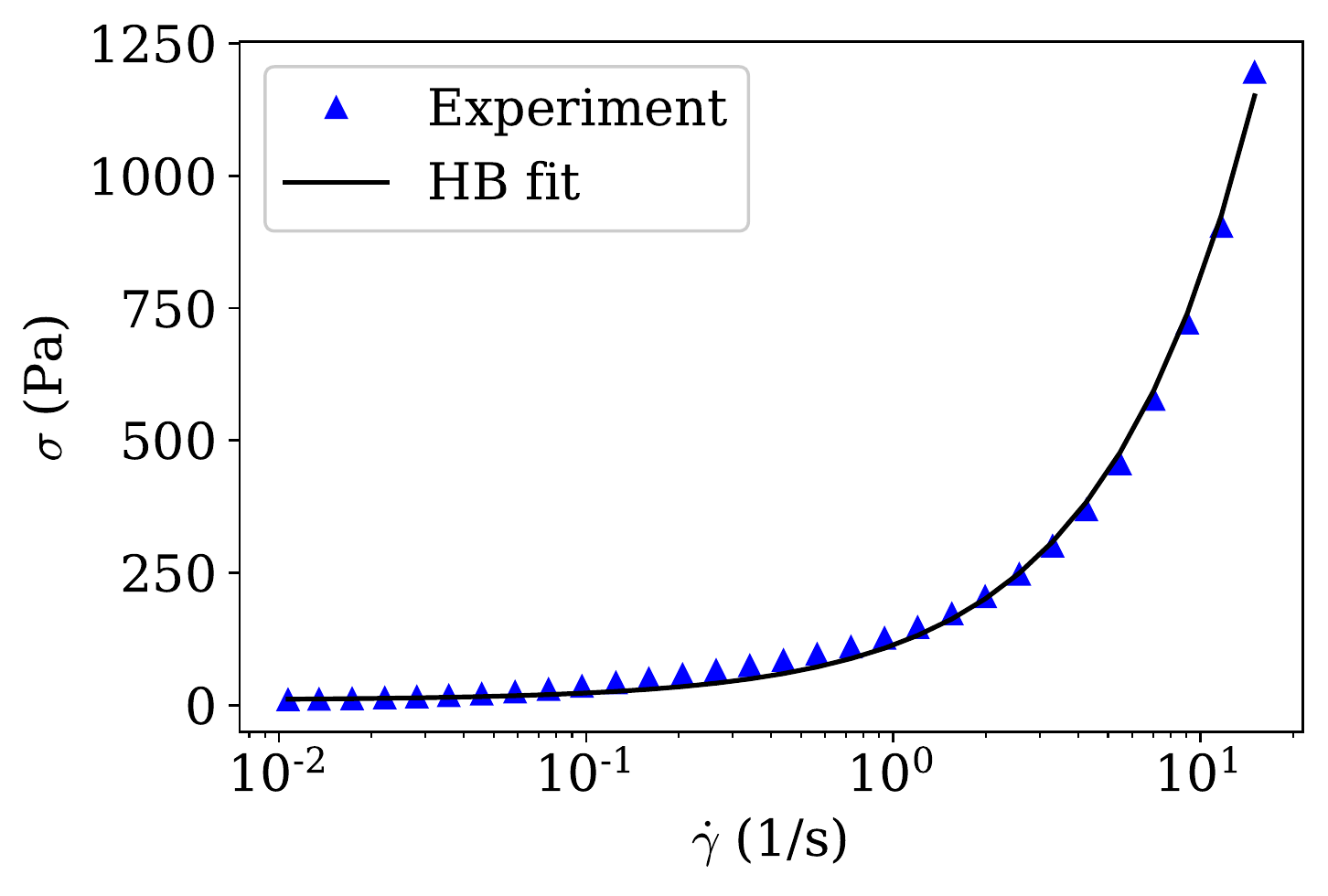}
    \caption{Flow curve data for DOWSIL TC-5550 thermal grease plotted on a semi-log plot. The yield stress $\sigma_y=9.5~\si{\pascal}$ is obtained by extrapolating the curve to $\dot\gamma=0~\si{s^{-1}}$, and  $k=104.0~\si{\pascal \second\tothe{n}}$ and $n=0.88$ are subsequently obtained by fitting to an HB model: $\sigma(\dot{\gamma})=\sigma_y + k \dot\gamma^n$.}
    \label{fig::DOW1_flowcurve_val}
\end{figure}

We performed an additional steady flow curve experiment by ramping down $\dot\gamma$ from $15~\si{\per\second}$ to $0.01~\si{\per\second}$ as shown in Fig.~\ref{fig::DOW1_flowcurve_val}. To characterize the steady flow curve, each data point was taken after $10~\si{\second}$ elapsed to ensure that a steady state was reached at that particular $\dot\gamma$. Next, we extrapolated the curve to $\dot\gamma=0~\si{\per\second}$ to obtain $\sigma_y$ \citep{Dinkgreve2017EverythingFluids}. With this value of $\sigma_y$ in hand, a steady HB model, namely $\sigma=\sigma_y + k \dot\gamma^n$, was fit to the flow curve data. Note that extrapolating $\sigma_y$ is preferred, as fitting this value simultaneously with the remaining HB parameters is quite sensitive to the number of data points used.

The HB fit of the steady flow curve serves as an alternative way of obtaining the parameters $\sigma_y$, $k$, and $n$ pertaining to DOWSIL TC-5550. Comparing the latter to the values inferred by the PINN serves as a ``sanity check'' on the learned unknown model parameters and their orders of magnitude. Recalling Table~\ref{tab::dow1}, the yield stress the PINN learns is $\sigma_y\approx 6.9~\si{\pascal}$, compared to $\sigma_y \approx 9.5~\si{\pascal}$ from extrapolating the flow curve; respectively, the PINN learns $n\approx0.85$, while the 95\% confidence interval  obtained from the HB fit is $n\in[0.85,0.92]$. These values, found by different methods, are in agreement. On the other hand, the consistency index learned by the PINN is $k\approx 214.2~\si{\pascal \second \tothe{n}}$, while the 95\% confidence interval obtained from the HB fit is $k\in[95.87,112.2]~\si{\pascal \second \tothe{n}}$. The $k$ values are not as close numerically (as the $\sigma_y$ and $n$ values), but it is generally expected that the consistency index $k$ of an HB fit can have a broad range, depending on the method of estimation \citep{Dinkgreve2017EverythingFluids}. Overall, this comparison gives us further confidence that our rheological characterization of the DOWSIL TC-5550 thermal grease in the stress buildup regime is consistent with its corresponding steady flow curve.

\section{Conclusion}
\label{sec::Conclusion}

Thermal greases are complex soft materials. The thermal performance and long-term degradation (pump-out and dry-out) of these greases, used as thermal interface materials within an electronic package, are strongly dependent on their rheological behavior. Previous studies of the rheology of thermal greases characterized only the steady-state flow curve (via the Herschel--Bulkley and Bingham models) \citep{Prasher2002RheologicalExperimental,Prasher2002RheologicalModeling}. In this work, we experimentally demonstrated the transient rheological behavior of thermal greases. We explained this observation on the basis that the shear-induced microstructural rearrangements of the high-thermal-conductivity filler particles dispersed within the grease's polymer matrix cause the time-dependent rheology. To understand this transient behavior from a fundamental point of view, we turned to a standard thixo-elasto-visco-plastic (TEVP) model \citep[e.g.,][]{Mahmoudabadbozchelou2021Rheology-InformedFluids} and a recently proposed nonlinear-elasto-visco-plastic (NEVP) model \citep{Kamani2021UnificationFluids}, and incorporated them in a data-driven machine learning framework. 

Specifically, we characterized the behavior of two commercial thermal greases: DOWSIL TC-5622 and DOWSIL TC-5550. The two thermal greases exhibit distinctive transient rheological behaviors---stress relaxation for DOWSIL TC-5622 and stress buildup for DOWSIL TC-5550,---which can be characterized by the TEVP and NEVP constitutive models, respectively. Startup flow protocol experiments at a constant shear rate were used to train a physics-informed neural network (PINN) and thus determine (``learn'') the unknown rheological model parameters in a data-driven manner. We performed startup flow experiments for a range of shear rates expected to be relevant to thermal applications for DOWSIL TC-5622 in the stress relaxation regime and for DOWSIL TC-5550 in the stress buildup regime, which served as the training data for developing the PINNs. The PINN's predictive ability was evaluated on ``unseen'' startup flow test data. The learned model parameters of DOWSIL TC-5622 were further validated by showing that the model's prediction of the stress evolution for an experiment with an input step-strain shear rate profile agrees with the experiments. Meanwhile, for DOWSIL TC-5550, the predicted steady-state model parameters showed good agreement with those obtained from a Herschel--Bulkley fit of the experimental flow curve.

Further, it is important to note that the characterized rheology (i.e., the learned model parameters) is only valid in the given flow regime of interest and may not hold outside of it. Indeed, the flow behavior of these complex soft materials, which further exhibit a yield stress, may not even be described by the models used herein in a different flow regime \citep{Bonn2017YieldMatter}. Of course, this limitation is a standard caveat of any rheological characterization. 

Nevertheless, having demonstrated and quantified the rheological regimes of both stress relaxation and stress buildup in two commercial thermal greases, beyond their standard steady-state characterization as shear-thinning fluids, we hope to have provided motivation for the need to develop more comprehensive constitutive models that can capture thermal grease behavior across multiple rheological regimes. In this work, we only reported average values of learned model parameters. An important avenue of future work would be to solve a Bayesian parameter inference problem using probabilistic machine learning computational tools, such as PyMC3 \citep{Salvatier2016ProbabilisticPyMC3}, with the constitutive models from this work, to obtain the distribution (rather than a single value) of the unknown model parameters. In doing so, the uncertainty in the learned model parameter would be characterized. We expect that the present results and future fundamental rheological studies of thermal greases will eventually yield a mechanistic understanding of why thermal interface materials may (or may not) resist common degradation mechanisms such as pump-out and dry-out.

\section*{Acknowledgements}
This work was supported by members of the Cooling Technologies Research Center (CTRC), a graduated National Science Foundation Industry/University Cooperative Research Center at Purdue University. 
We further thank Dr.\ Andres Becerra and Dr.\ Danielle Berry from the DOW Chemical Company for providing us with thermal greases and having insightful discussions regarding the rheology of DOWSIL TC-5622 and DOWSIL TC-5550 thermal greases.


\section*{Data Availability}
The data that support the findings of this study (rheological experiments and corresponding Python notebooks, within which PINNs are implemented) are openly available in a repository archived at  \url{https://dx.doi.org/10.5281/zenodo.8303195}.

\bibliography{PINN_references_ICC.bib}


\appendix

\section{Extrapolation abilities of the PINNs}
\label{app:PINN_extrapolation}
In this appendix, we illustrate the prediction of stress relaxation profiles in Fig.~\ref{fig::DOW5622_expt} at shear rates outside the range of training data-set shear rates, i.e., we test the extrapolation powers of the presented PINN formulation for DOWSIL TC-5622 thermal grease. Specifically, we train the PINN on stress relaxation data with $\dot\gamma = 2$ to $5~\si{\per\second}$ and test the trained model on $\dot\gamma=7~\si{\per\second}$ and $\dot\gamma=9~\si{\per\second}$ which are outside the interval of training $\dot\gamma$ values. To facilitate testing/prediction for shear rates outside the training range, we must also consider the initial condition (IC) residual and add another term to the loss function in Eq.~\eqref{eq:OverallLoss}. Specifically, now
\begin{subequations}
\begin{equation} 
    L = w_\mathrm{data} \mathrm{MSE_\mathrm{data}} + w_{\sigma} \mathrm{MSE}_{\sigma} + w_{\lambda} \mathrm{MSE}_{\lambda} + w_\mathrm{ic} \mathrm{MSE}_\mathrm{ic},\label{eq:LossDef1}
\end{equation}
where
\begin{equation}
    \mathrm{MSE_\mathrm{ic}} = \frac{1}{P} \sum_{i=1}^{P} | \sigma(t=0;\dot\gamma_i) - \sigma_{\mathrm{ic,fit}}(\dot\gamma_i) |^2.
\end{equation}
\label{eq:OverallLoss1}%
\end{subequations}
Here, $w_\mathrm{ic}=1$ is weight for the IC residual term, $\mathrm{MSE}_\mathrm{ic}$ is the mean squared error in the IC residual, $P=50$ is the number of $\dot{\gamma}$ data points sampled, and 
\begin{equation}
    \sigma_\mathrm{ic,fit}(\dot\gamma) = (0.6491~\si{\pascal\second})\dot{\gamma} + 0.3045~\si{\pascal}
    \label{eq:sigma_ic_fit}
\end{equation}
is the experimental fit of the initial condition, $\sigma(t=0$), as a function of the given startup shear rate $\dot\gamma$. 

\begin{figure*}[ht]
    \centering
    \begin{subfigure}[t]{0.49\linewidth}
      \includegraphics[width=\linewidth]{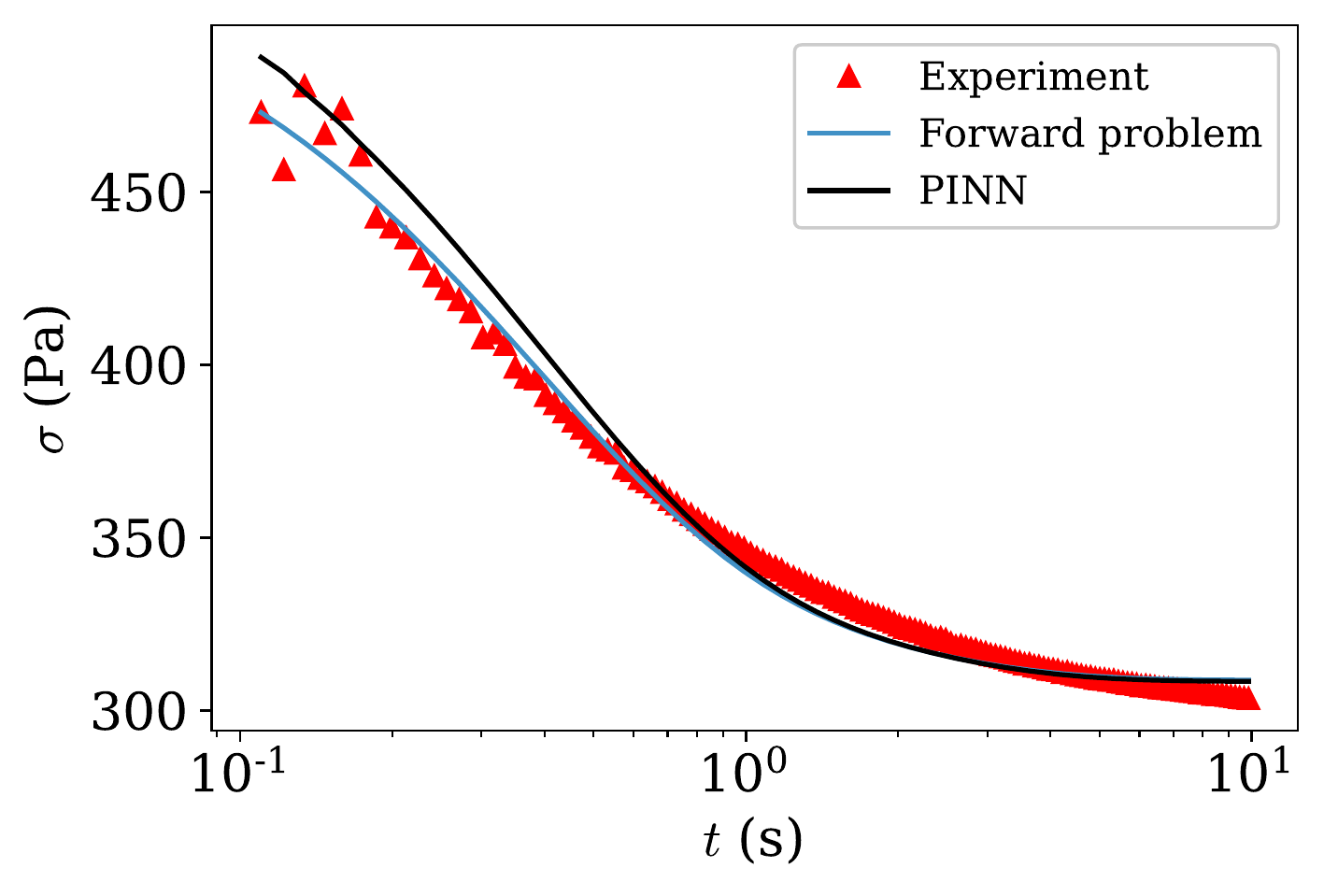}  
      \caption{}
      \label{fig:5622_extrapolate_7s}
    \end{subfigure}
    \begin{subfigure}[t]{0.49\linewidth}
      \includegraphics[width=\linewidth]{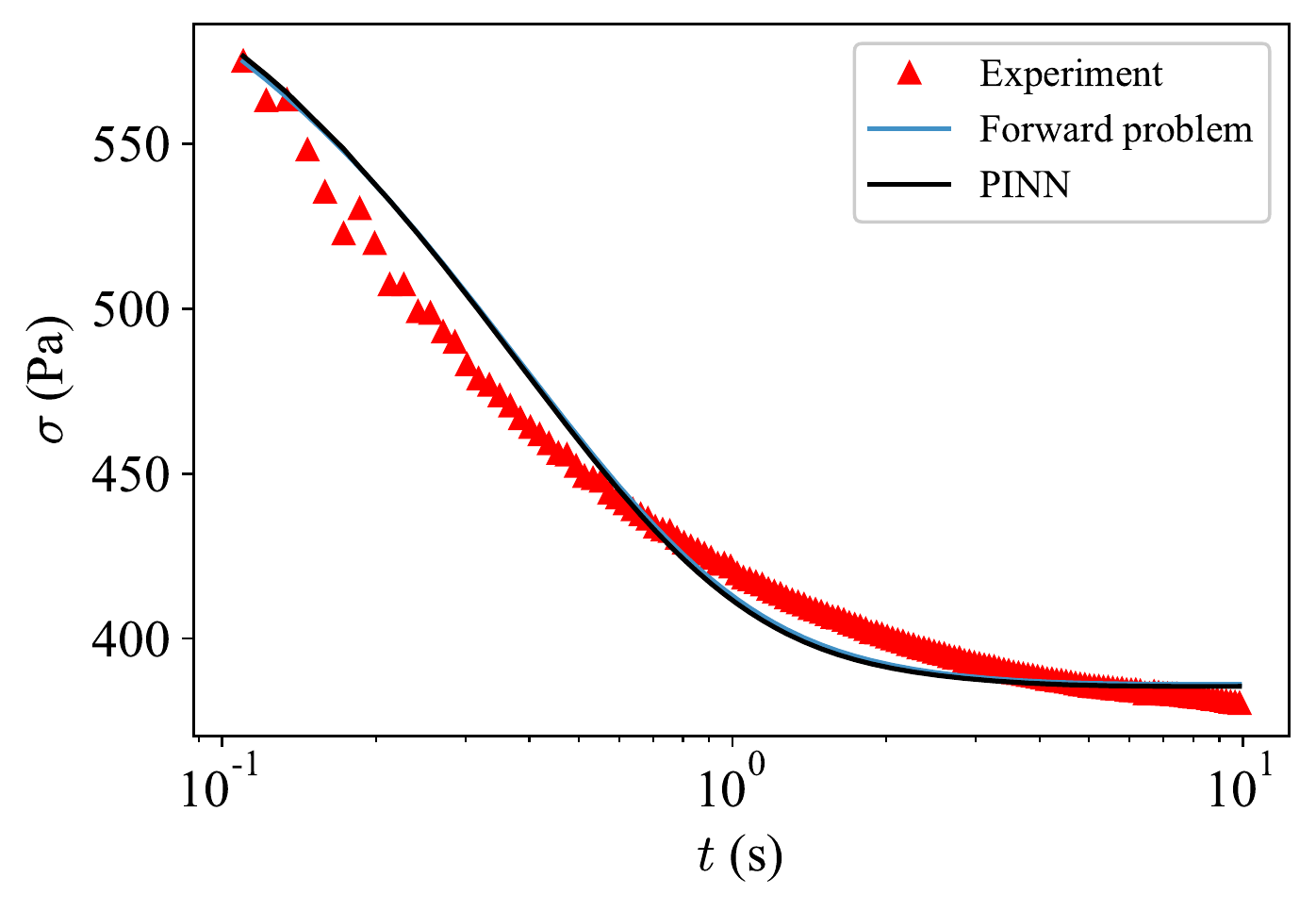}  
      \caption{}
      \label{fig:5622_extrapolate_9s}
    \end{subfigure}
    \caption{Comparison between a PINN prediction that requires extrapolation and the experimental data, for (a) $\dot\gamma = 7~\si{\per\second}$ and (b) $\dot\gamma = 9~\si{\per\second}$, for the DOWSIL TC-5622 thermal grease. The PINN was trained on the experimental data for $\dot\gamma = 2$ to $5~\si{\per\second}$.}
    \label{fig::dow5622_extrapolate}
\end{figure*}

We tested predictions at $\dot\gamma=7~\si{\per\second}$ and $\dot\gamma=9~\si{\per\second}$ after training for $1.5\times10^6$ epochs to achieve a loss $L\le 6 \times 10^{-4}$.
As shown in Fig.~\ref{fig::dow5622_extrapolate}, the PINN is successful in predicting the stress profiles at $\dot\gamma=7~\si{\per\second}$ and $\dot\gamma=9~\si{\per\second}$ (when trained on $\dot\gamma = 2$ to $5~\si{\per\second}$) \emph{if} the initial condition $\sigma(t=0)$ is provided for each $\dot\gamma$. (The case without an initial condition is not shown, as the agreement is poor.) Indeed, one cannot expect the PINN to be able to guess this initial condition, which is a function of the initial state of the thermal grease in the rheometer, and it should indeed be supplied to the PINN (or the forward problem solver) for successful extrapolation. 

In Fig.~\ref{fig::dow5622_extrapolate}, we also show the result of solving Eqs.~\eqref{eq:TEVP-dim} numerically as a forward problem. In this case, we use the constitutive model parameters learned by the PINN and provide an appropriate IC based on Eq.~\eqref{eq:sigma_ic_fit}.

\section{Characterizing the yield stress of DOWSIL TC-5622 via a flow curve experiment}
\label{app::dow5622-flow-curve}

\begin{figure}[ht]
    \centering
    \includegraphics[width=\linewidth]{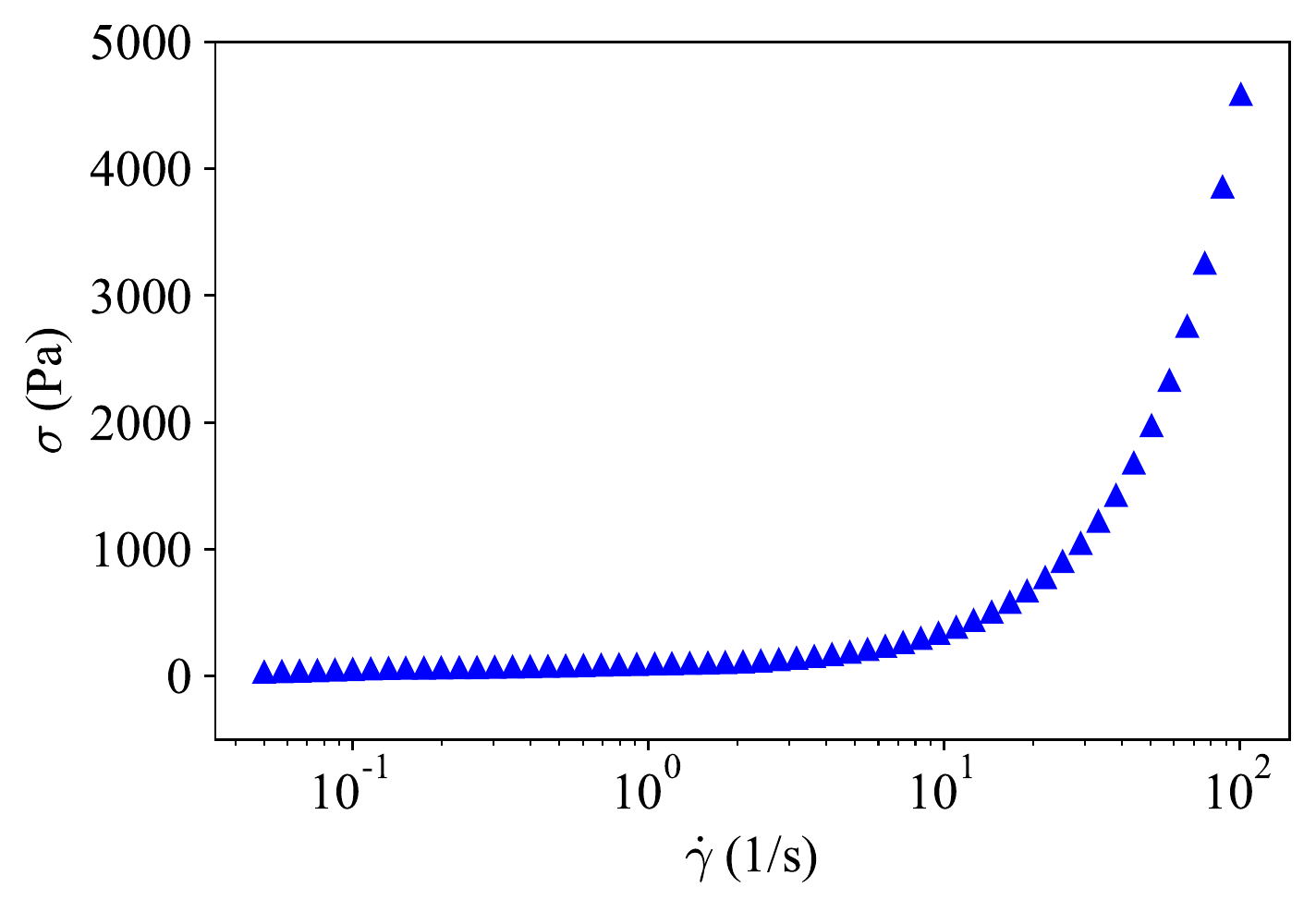}
    \caption{Flow curve data for the DOWSIL TC-5622 thermal grease plotted on a semi-log plot. A yield stress of $\sigma_y=29.5~\si{\pascal}$ is obtained by extrapolating the curve to $\dot\gamma=0~\si{\per\second}$.}
    \label{fig::DOW-5622-fc}
\end{figure}

We performed a flow curve experiment to determine $\sigma_y$ for DOWSIL TC-5622 to enable an additional check on the value ($\sigma_y\approx 31.1~\si{\pascal}$) given in Table~\ref{tab::dow5622} obtained via physics-informed machine learning. Figure~\ref{fig::DOW-5622-fc} shows the experimental flow curve for DOWSIL TC-5622. We obtain $\sigma_y\approx29.5~\si{\pascal}$ from the data in Fig.~\ref{fig::DOW-5622-fc} by extrapolating the ramp-down cycle of the flow curve to $\dot\gamma=0~\si{\per\second}$ \citep{Dinkgreve2017EverythingFluids}. (Again, note that we extrapolate the flow curve to $\dot\gamma=0~\si{\per\second}$ to obtain $\sigma_y$, rather than obtaining it from an HB fit of the flow curve, because the HB fit's values are quite sensitive to the number of flow-curve data points used for fitting.) This comparison gives us further confidence that our rheological characterization of the DOWSIL TC-5622 thermal grease in the stress relaxation regime is consistent with its corresponding steady flow curve. 

\begin{table*}[ht]
    \setlength{\tabcolsep}{5pt}
    \begin{tabular}{l|ll|ll|ll}
      \hline
      \hline
      Parameter  & Initial guess 1 & Learned value & Initial guess 2 & Learned value & Initial guess 3 & Learned value \\
      \hline
      $G$ (\si{\pascal}) & $80.0$ & $190.0$ & $160.0$ & $203.0$ & $80.0$ & $192.0$ \\
      $\eta_s$ (\si{\pascal\second}) & $44.5$ &  $39.4$ & $100.0$ &  $39.3$ & $75.0$ &  $39.3$ \\
      $k_{+}$ (\si{\per \second}) & $1.0$ & $0.061$ & $0.5$ & $0.052$ & $0.0$ & $0.057$ \\
      $k_{-}$ (--) & $1.0$ & $0.064$ & $0.5$ & $0.061$ & $0.0$ & $0.059$ \\
      $\sigma_y$ (\si{\pascal}) & $2.0$  &  $30.5$ & $20.0$  &  $17.3$ & $10.0$  &  $20.5$ \\
      $\eta_p$ (\si{\pascal\second}) & $50.0$ & $30.8$ & $25.0$ & $37.4$ & $75.0$ & $32.1$ \\
      \hline
      \hline
    \end{tabular}
    \caption{Learned values of the parameters of the TEVP rheological model~\eqref{eq:TEVP-dim} for DOWSIL TC-5622, starting from three different sets of initial guesses.}
    \label{tab::dow5622-parspace}
\end{table*}

\section{Influence of initial guess on the learned unknown model parameters}
\label{app::parspace}

To ascertain the influence of initial guesses on the unknown model parameters, we trained the PINN for the stress relaxation regime (for DOWSIL TC-5622 thermal grease) with three different initial guesses of the unknown model parameters. We trained the PINN to a loss of $L \le 6\times10^{-4}$ for each set of initial guesses. As seen from Table~\ref{tab::dow5622-parspace}, the learned model parameters are close to each other despite starting the training from different initial guesses. Thus, we conclude that the results of our PINN algorithm are not sensitive to the initial guesses for the unknown model parameters.

\begin{figure*}[ht]
    \centering
    \begin{subfigure}[t]{0.49\linewidth}
      \includegraphics[width=\linewidth]{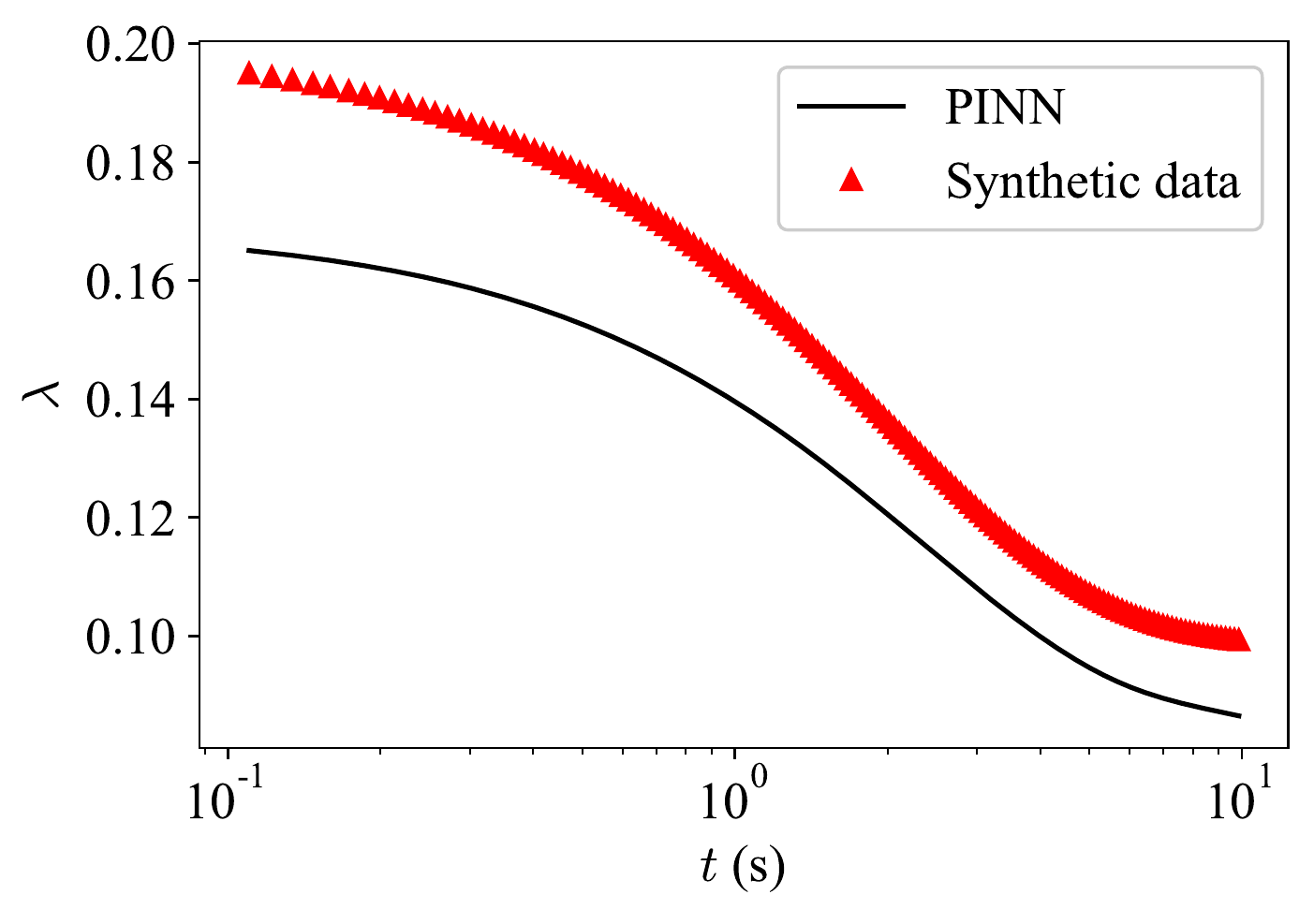}  
      \caption{}
    \end{subfigure}
    \begin{subfigure}[t]{0.49\linewidth}
      \includegraphics[width=\linewidth]{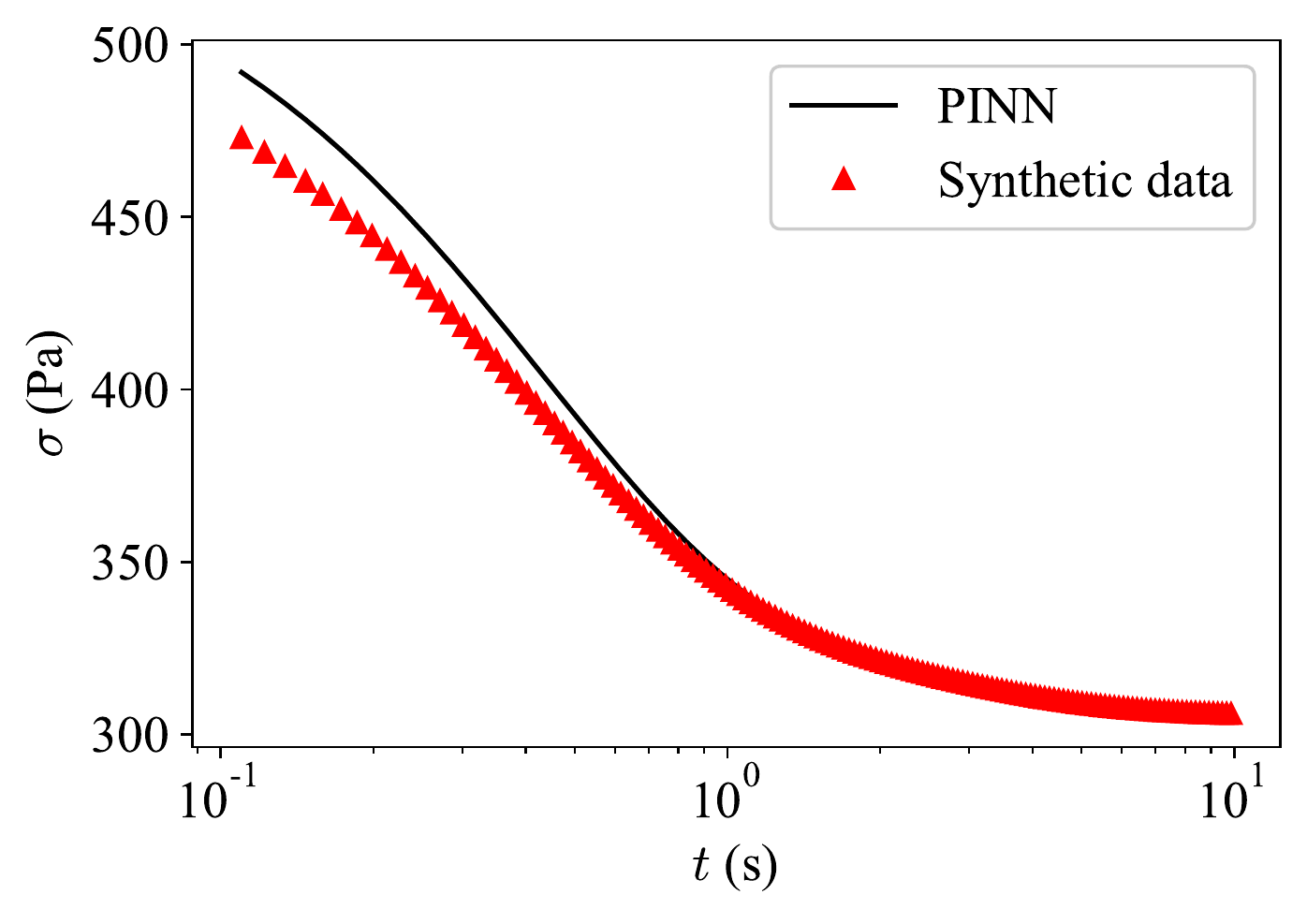}  
      \caption{}
    \end{subfigure}
    \caption{Testing the PINN on unseen data (for startup flow with $\dot\gamma = 7~\si{\per\second}$), having trained it on synthetic data generated by solving the TEVP model~\eqref{eq:TEVP-dim}, for $\dot\gamma=2$ to $10~\si{\per\second}$ skipping  $\dot\gamma=7~\si{\per\second}$, numerically as a forward problem with the model parameter values given in Table~\ref{tab::TEVP_synthetic}. The synthetic $\lambda(t)$ data is not provided to the PINN during training.}
    \label{fig::TEVP_synthetic}
\end{figure*}

\section{TEVP inverse problem solution via a PINN trained on synthetic data}
\label{app:PINN_synthetic_data}

In this appendix, to demonstrate the robustness of the inferred unknown model parameters and highlight PINNs' ability to ``learn'' the microstructure evolution for which no data is provided, we generated synthetic data by solving Eqs.~\eqref{eq:TEVP-dim} numerically as a forward problem with the model parameters values from Table~\ref{tab::TEVP_synthetic}, and given initial values for the stress, $\sigma(t=0)$, and microstructure $\lambda(t=0)$. Then, we trained a PINN based on this synthetic stress data (for $\dot\gamma=2$ to $10~\si{\per\second}$, skipping  $\dot\gamma=7~\si{\per\second}$) to solve the inverse problem and obtain ``new'' predictions for $\sigma(t)$, $\lambda(t)$, and the six model parameters. Note that, in training the PINN, we do not use the $\lambda(t)$ synthetic data since this data is also not available from the experiment.

\begin{table}[b]
    \setlength{\tabcolsep}{10pt}
    \begin{tabular}{lll}
      \hline
      \hline
      Parameter  & For synthetic data & From PINN \\
      \hline
      $G$ (\si{\pascal}) & $213.0$ & $201.0$ \\
      $\eta_s$ (\si{\pascal\second}) & $39.2$ & $39.6$ \\
      $k_{+}$ (\si{\per \second}) & $0.049$ & $0.037$\\
      $k_{-}$  & $0.064$ & $0.057$\\
      $\sigma_y$ (\si{\pascal}) & $19.6$ & $23.1$\\
      $\eta_p$ (\si{\pascal\second}) & $42.5$ & $43.6$\\
      \hline
      \hline      
    \end{tabular}
    \caption{Model parameters used to generate synthetic TEVP models stress data and the corresponding values inferred by the PINN trained on this synthetic data.}
    \label{tab::TEVP_synthetic}
\end{table}

Table~\ref{tab::TEVP_synthetic} shows that the unknown model parameters inferred from the synthetic data compare favorably with the values used to generate this data, indicating that the PINN's inference of these values is robust and accurate. Figure~\ref{fig::TEVP_synthetic} shows a comparison of the PINN predictions for the test (unseen) data at $\dot\gamma=7~\si{\per\second}$ to the synthetic data itself. We observe that there is some systematic deviation between the PINN prediction and the synthetic $\lambda(t)$ data. However, even though the data for the structure parameter was not used in training the PINN, the PINN is able to capture the monotonically decreasing trend 
indicative of the microstructure breakdown. Finally, there is a small deviation between the PINN prediction and the synthetic data for the stress at early times ($t<1~\si{\second}$). This small discrepancy is overexaggerated by the semi-log plotting scale.

\end{document}